\documentclass[preprint,12pt]{elsarticle}

\usepackage{amssymb}
\usepackage{amsmath}
\usepackage{color}
\usepackage{epsfig,graphics,graphicx,subfigure}

\journal{Physica A}

\begin{document}

\begin{frontmatter}

\title{Mass media and heterogeneous bounds of confidence in continuous opinion dynamics}
 \author[label1,label2]{M.~Pineda}
 \author[label2]{G.~M.~Buend\'{\i}a}
 \address[label1]{Department of Mathematics, University of Dundee, Dundee DD1 4HN, United Kingdom}
 \address[label2]{Department of Physics, Universidad Sim\'on Bol\'{\i}var, Caracas 1080, Venezuela}

\begin{abstract}
This work focus on the effects of an external mass media on continuous opinion 
dynamics with heterogeneous bounds of confidence.
We modified the original Deffuant et al. and Hegselmann and Krause models to 
incorporate both, an external mass media and a heterogeneous
distribution of confidence levels. We analysed two cases, one where only two 
bounds of confidence are taken into account, and other were each individual of 
the system has her/his own characteristic level of confidence. We found that, in the absence
of mass media, diversity of bounds of confidence can improve the capacity of the systems to 
reach consensus. We show that the persuasion capacity of the external message is optimal for 
intermediate levels of heterogeneity. Our simulations also show the existence, for certain parameter values, 
of a counter-intuitive effect in which the persuasion capacity of the mass 
media decreases if the mass
media intensity is too large. We discuss similarities and differences between 
the two heterogeneous versions
of these continuous opinion dynamic models under the influence of mass media.  
\end{abstract}


\end{frontmatter}

\section{Introduction}
\label{introduction}
The physics of social systems deals with the application of methods from physics 
to understand the complex mechanisms behind social phenomena. Perhaps one of the 
most intriguing challenges of this popular topic of research is to explain the 
development of consensus on some issue out of social interactions, despite 
that initially all the individuals had different 
opinions. The usual approach is to set up stylised dynamical models in which 
properly quantified individual's opinions evolve according to realistic enough 
communication rules, that define how individuals influence and are influenced by 
other members of the society \cite{socio,castellano,stauffer}.

Recently, Deffuant et al. (DW) and Hegselmann and Krause (HK) 
developed a set of models in which opinions are represented by continuous 
varying quantities \cite{deffuant0,krause0,lorenz1}. To mimic social interactions, both models 
implement the so-called bounded confidence mechanism by which two individuals 
only influence each other if their opinions differ by less than some 
amount. Another common ingredient is an agreement mechanism, by which 
individuals that overcome the bounded confidence condition adjust their opinion 
towards an average value. In the DW model, a process similar to the collision of 
two molecules or atoms in the kinetic theory of gases is considered. The 
opinion of the two individuals changes simultaneously, and there is an extra 
parameter that controls how fast the opinion converge. This model describes 
situations where the interaction between individuals occurs face to face. In the 
HK model, there is not convergence parameter, and the interaction is 
not longer related with an scattering process, because the individuals 
simultaneously change their opinions to the average opinion of all other 
individual that satisfied the bounded confidence condition (a global 
communication takes place in large groups). These models of continuous opinion 
dynamics under bounded confidence may be very useful to analyse cases where one 
has to respond to a single issue (rating a politician or a product, for 
example) with a real number which can vary continuously in a certain range. It 
is already well established that there are different critical confidence levels 
above which a state of consensus is always reached. Below these consensus 
thresholds, the population splits into two (polarization) or more 
(fragmentation) sets of non-interacting clusters with the same opinion in each 
of them \cite{lorenz1,naim}. However, it has been also recognised that such 
uniform and noninteracting states are not very realistic, this being one of the 
reasons why several interesting modifications to these models have been recently 
introduced \cite{socio,castellano,pineda0,jensen1,jensen2}.

Most of the modifications assume situations where all individuals in a given 
society have the same level of confidence. However, it is clear that due to many 
complex psychological and physiological factors, a more realistic 
assumption is to consider systems where 
individuals are allowed to have different bounds of confidence. Surprisingly, it 
was found, using an interactive Markov chain formulation, that it is possible 
to reach spontaneous ordered states (consensus) 
in the HK model and DW model with two different bounds of confidence, but not 
for the corresponding homogeneous cases \cite{lorenz2}. More recently, a HK 
type model, where individuals are divided into not just two but several groups, 
with different bounded confidence levels, was also analyzed. In this case, the 
number of opinion clusters increases with the number of individuals who have a 
very low confidence level \cite{kou}. Several other interesting results about 
heterogeneous bounds of confidence have also been reported in recent years 
\cite{bullo,deffuant3,shang,liang}.

The dynamic of opinions in real societies is also affected by many exogenous 
factors, being the mass media one of the most important (T.V., blogs, 
newspapers, etc). In this regard and with the aim to make models more realistic, 
some works study the conditions for an efficient spreading of propaganda in 
DW type models where homogeneous individuals (equal level of confidence) 
interact with their neighbours and with the mass media using a bounded 
confidence mechanism \cite{pineda,carletti,walter}. Some of these works have
 reported the very counter-intuitive observation 
that the system can spontaneously order in a state different from the imposed by 
the mass media, in contradiction with what it is found in classical physics 
where spin systems monotonously align towards strong external applied fields. 
It was found that this situation could arise in other non-equilibrium 
models provided they allow for non-interacting states \cite{avella}. 

However, besides the modified models mentioned above, there are few studies 
that analyse the combined effect of mass media and heterogeneous 
bounds of confidence in continuous opinion dynamics \cite{gargiulo}. We 
believe that, given the strong global exchange of information and the strong 
presence of the media, it is worth to study in more detail the effect of both 
contributions \cite{bullo2}.  In order to shed some light on the emerging phenomena obtained 
after the combination of the endogenous and exogenous factors described above, 
we analyse extended versions of the DW model and the HK model in which 
individuals that have their own bound of confidence are allowed to interact 
with their neighbours and with an external imposed mass media. We want to 
analyse the impact of heterogeneous bounds of confidence in the adoption or 
rejection of an external message or mass media. We also analyse how the 
combination of these two factors affects the 
formation of consensus.  

In the next section we present the heterogeneous DW model in the presence of 
mass media. In Section~\ref{sec:HK}, we compare the results of the 
heterogeneous 
DW model with the heterogeneous HK model also under mass media. Summary and conclusions are presented in Section~\ref{sec:conclusion}.
\section{The heterogeneous DW model with mass media}
\label{sec:heteDW}
Usually, the effects of mass media and heterogeneous bounds of confidence
in the DW model are analysed separately. Very few cases consider those 
ingredients acting together. In this section, we consider a modification 
in which a heterogeneous distribution of confidence levels and a constant 
exogenous mass media are added to the original DW rules. 

To begin the analysis, lets start with a population composed 
by $N$ individuals in a fully connected network. The opinion $x^{i}_{n}$ on a 
given issue that individual $i$ has at time-step $n$ is a real variable in the 
interval $[0,1]$. We assign to each agent $i$  a constant bound of confidence level,  
$\epsilon_{i}$, that runs from $0$ to $1$. This control parameter reflects the 
willingness of an individual to change his opinions after interacting with others 
or with an external mass media.  In certain way, $\epsilon_{i}$, measures the 
level 
of tolerance of individual $i$. We also introduce an external mass media $S$ 
which in principle can take any value between $0$ and $1$. Finally, we 
assign to each individual a constant convergence parameter, $\mu_{i} \in 
[0,0.5]$. Initially, it is assumed that the values $x^{i}_{0}$ for each 
individual are randomly
distributed in the interval $[0,1]$. At 
time step $n$, an individual $i$ is selected at random:

1. With probability $m$, if $|x^{i}_{n}-S| < \epsilon_{i}$, individual $i$ interacts 
with the external field $S$ such that,  
\begin{equation}
x^{i}_{n+1}=x^{i}_{n}+\mu_{i}(S-x^{i}_{n});
\end{equation}
if $|x^{i}_{n}-S| \ge \epsilon_{i}$ the opinion of the individual does not change.  
In any case, the cycle starts again.

2. Otherwise, with probability $1-m$, a different individual $j$ is selected at 
random: 
if $|x^{i}_{n}-x^{j}_{n}| < \epsilon_{i}$, then
\begin{equation}
x^{i}_{n+1}=x^{i}_{n}+\mu_{i}(x^{j}_{n}-x^{i}_{n});
\end{equation}
also if $| x^{i}_{n}-x^{j}_{n}| < \epsilon_{j}$, individual $j$ also changes its opinion 
according with
\begin{equation}
x^{j}_{n+1}=x^{j}_{n}+\mu_{j}(x^{i}_{n}-x^{j}_{n});
\end{equation}
if $| x^{i}_{n}-x^{j}_{n}| \ge \epsilon_{i}$ opinion of individual $i$ does 
not change. In the same way, if  $| x^{i}_{n}-x^{j}_{n}| \ge \epsilon_{j}$, 
the opinion of individual $j$ does not change. In any case, the cycle 
starts again.

According with these dynamical rules, when individuals interact between them or 
with the mass media, they can change their opinion depending on 
their particular bounded confidence levels. Whether the opinions changed 
or not, time
increases $n \rightarrow n+1$. We introduce the time variable $t=n\Delta t$, 
$\Delta t=1/N$,  where the time unit is  given in terms of Monte Carlo steps ($MCS$). 
In our simulations, we set the mass media "opinion", $S=1$. The probability $m \in [0,1]$ 
represents the strength of this mass media or external field compared with the 
strength of the peer interaction. In this work, we only analyse cases where 
$\mu_{i}=\mu_{j}=\mu=0.5$.

\subsection {A society of individuals with two bounds of confidence}
\label{subsec:twogroups}
We start our analysis with only two bounds of 
confidence, $\epsilon_{1}$ and $\epsilon_{2}$. We analyse the opinion evolution 
of a system composed by two types of individuals that differ in the way they 
relate with each other and with an 
external message. Individuals with the lower confidence bound constitute the closed-minded 
group, and the ones with the higher confidence bound are the open-minded group. To simplify the analysis, we will assume that the two 
subsystems have the same number of individuals ($N_{1}=N_{2}=N/2$). 

\subsubsection{Behavior in the absence of mass media ($m=0$)}
\label{subsubsec:nomass}
\begin{figure}
\begin{center}
\mbox{\includegraphics[clip,angle=270,width=.6 \textwidth]{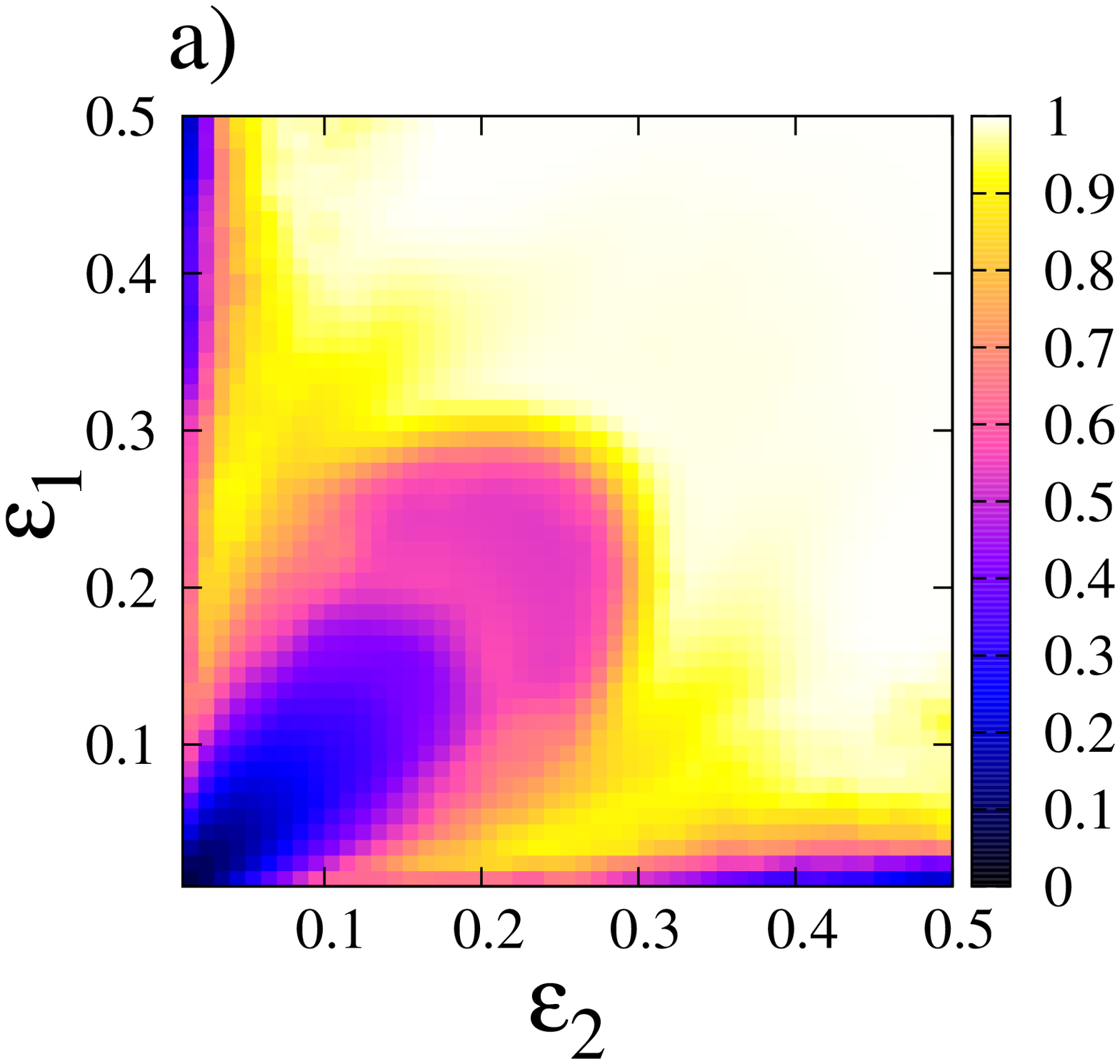}}
\mbox{\includegraphics[clip,angle=270,width=.6 \textwidth]{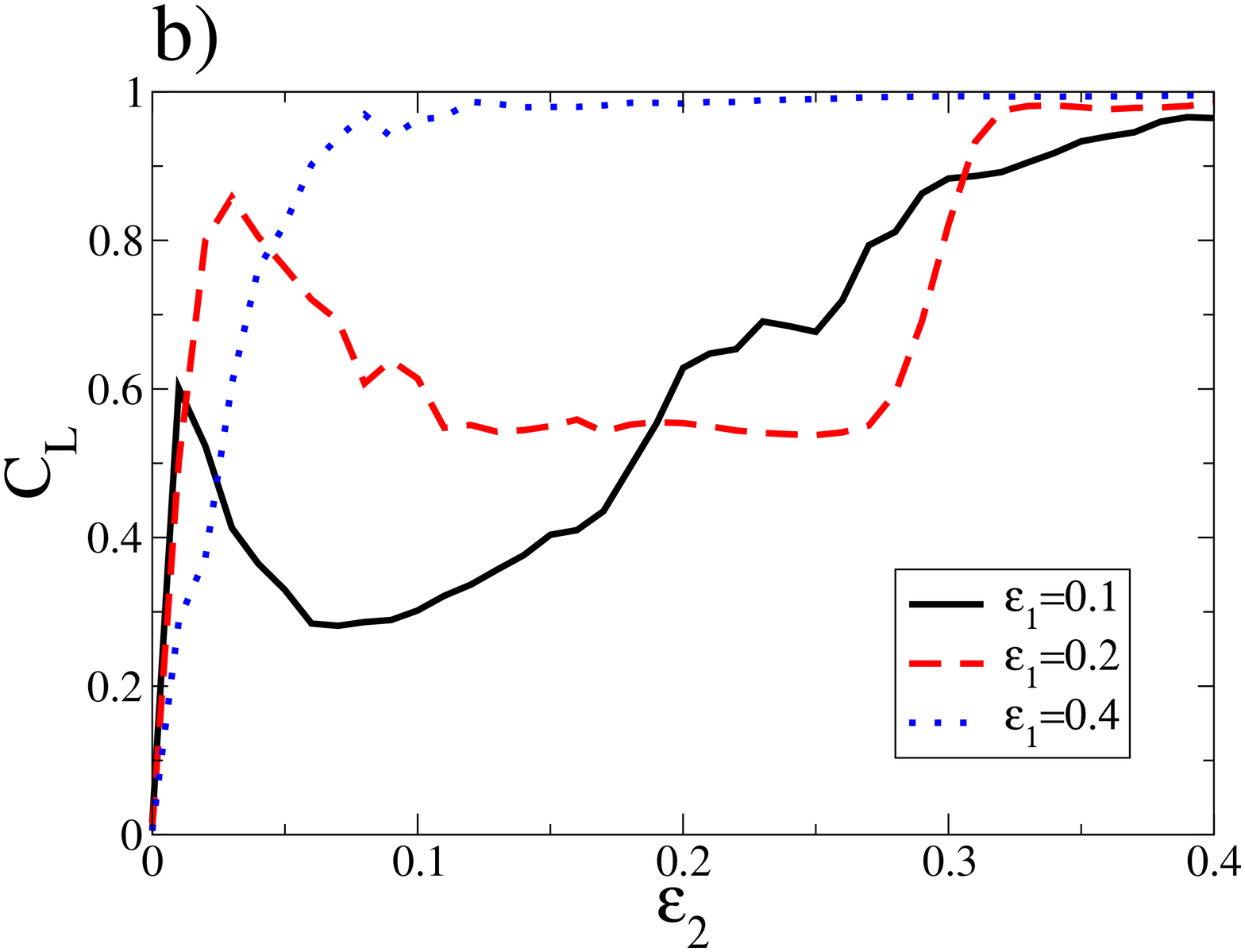}}
\caption{
Heterogeneous DW model with two bounds of confidence ($\epsilon_{1}$ and $\epsilon_{2}$) and $m=0$.
 a) Density plot of the normalised average size of the largest domain, $C_L$, as a function of bounded
confidence parameters $\epsilon_{2}$ and $\epsilon_{1}$. b) The same order parameter, 
$C_L$, as a function of $\epsilon_{2}$ for $\epsilon_{1}=0.1$ (black solid line), $0.2$ (red-dashed line), 
and $0.4$ (blue-dotted line).  In all figures the total number of individuals is $N=1000$ with each subsystem
 having $N/2$ individuals (the system is  symmetric under the exchange of confidence levels). Each data point is plotted at
 $2 \times 10^{5}$ MCS after averaging over $100$ runs starting from uniformly distributed initial opinions inside the interval $[0,1]$.  
 }\label{fig1}
\end{center}
\end{figure}

For comparison purposes we start studying the behavior of a system of 
heterogeneous individuals in the absence of any external signal, $m=0$ 
\cite{lorenz2}. We are going to measure the  normalised average size of the largest domain, 
$C_L=N_L/N$, as a function 
of $\epsilon_{1}$ and $\epsilon_{2}$. Then $C_L$ plays the role of an order 
parameter. The quantities $N_{L}$ and $N$ are 
the number of individuals in the largest domain and 
the total number of individuals in the system, respectively. $C_L$ runs from 
$0$ to $1$. 
If $C_{L}=1$, the largest domain contains the whole population; the population 
reaches a total consensus. In the opposite case, if it is
zero, the population is in a disordered state where no particular opinion 
group dominates. Figure~\ref{fig1}a) shows a density
plot of $C_{L}$ in the plane ($\epsilon_{2}$, $\epsilon_{1}$), for $N=1000$.
The phase space must be symmetric under exchange of $\epsilon_{1}$ and 
$\epsilon_{2}$. The few differences between both sides of the diagonal, 
 are due to finite size effects.  
The diagonal corresponds to the  limit case in which the 
original DW model is recovered ($\epsilon_{1}=\epsilon_{2}=\epsilon$). 
For example, 
one can verify the well known fact that, at $\epsilon_{c} \approx 0.27$ a sharp 
transition from a polarized state with two almost equally sized big clusters 
($C_L \approx 0.5$) to a state of consensus ($C_L \approx 1$), occurs 
(again, any minor discrepancy with the actual value 
is mainly due to finite size effects) \cite{lorenz1} . Still on the diagonal, 
as  
$\epsilon$ decreases, there are regions composed by more and more small opinion 
clusters separated by distances larger than $\epsilon$, which are delimited by 
other bifurcation points. When $\epsilon=0$, the expected result, 
$C_{L}=0$, is obtained. Similarly, the limit in which one of the groups always 
has a very small bounded confidence level (close to the axes) is characterised 
by a totally 
disordered state. Under these conditions, the open-minded 
individuals are always interacting with a noisy environment generated by the 
initial random distribution of closed-minded individuals. 

However, the most relevant finding presented in Fig.~\ref{fig1}a) is the 
existence of regions outside the diagonal line with very large values of 
$C_{L}$. In particular for values of $\epsilon_{1}$ and $\epsilon_{2}$ way 
smaller than the critical point of the 
homogeneous case $\epsilon_{c} \approx 0.27$.  This surprising finding, first 
reported in a Markov chain implementation of this model \cite{lorenz2}, is better 
observed in Fig.~\ref{fig1}b), where $C_{L}$ is plotted as a function of 
$\epsilon_{2}$, for three characteristic values of $\epsilon_{1}$.
However, given the symmetry of the model, it is enough to analyse
 what happen when $\epsilon_{2}$ runs  $0$ to $\epsilon_{1}$.  For $\epsilon_{1}=0.1$
 and $0.2$, it is clear that a maximum of $C_{L}$ appears for intermediate values of $\epsilon_{2}$, meaning that the 
heterogeneous system ($\epsilon_{2}<\epsilon_{1}$) may have spontaneous states 
with higher levels of order (high levels of consensus) than that of the 
homogeneous case ($\epsilon_{2}=\epsilon_{1}$). For the parameters
selected in this figure, the highest spontaneous ordering is characterised by 
 a largest domain that contains more than half of the population 
($C_{L}>0.5$). For the larger value, $\epsilon_{1}=0.4$, the phenomenon 
disappears and 
the whole system just moves from a disordered state ($C_{L} \approx 0$) to an ordered 
state characterised by almost total consensus ($C_{L} \approx 1$).

In this model where the individuals have different bounds of 
confidence, the consensus state that emerges from an uniform initial 
distribution of opinions inside the interval $[0,1]$, might not lie in the center of that interval. 
Instead, a drifting phenomenon occurs in which the final consensual state moves 
towards one of the two extremes
of the opinion space \cite{lorenz1,naim,pineda0,lorenz2} (the initial average opinion of approximately $1/2$ is not necessary conserved). 

To obtain some insight into the process that drives the formation of 
these consensual states, we include an appendix where we perform an analysis of 
the master equation of the system.
 The main conclusion
drawn from this study is that a consensus state may indeed emerge
for values of confidence levels below the consensus threshold of the homogeneous
case. As an example, in Fig.~\ref{fig10} we show the time evolution of the 
probability distribution of opinions, obtained by solving a set of coupled 
master equations corresponding a system where half the individuals have 
$\epsilon_{1}=0.22$ and the other half, $\epsilon_{2}=0.1$  
Both confidence values are far below than $\epsilon_{1}=\epsilon_{2} \approx 0.27$.  
According with the rough $\frac{1}{2\epsilon}$-rule reported in \cite{deffuant0,lorenz1},
which postulates that the number of big opinion clusters after a steady 
state is reached, are approximately
determined as the integer part of $\frac{1}{2\epsilon}$, two and five, big 
clusters must be formed
in the homogeneous cases with $\epsilon_{1}$ and $\epsilon_{2}$, 
respectively.
However, our analysis shows that a consensual state is reached when there is 
mixture
of both confidence levels.  An initial group of closed-minded individuals in the 
center of the
opinion space, pulls the open-minded individuals towards the center. 
Simultaneously, these
 open-minded individuals pull closed-minded individuals located on both 
extremes of $x=1/2$
 towards the central cluster of closed-minded individuals. This leads to the 
formation of a state of
 consensus (neglecting a small proportion of extremists).  In Fig.~\ref{fig11}, we also show that
clusters of individuals with low enough values of confidence levels, and with 
opinions very close to 
the extremes, can split the open-minded individuals toward the two extremes by 
forming two big clusters.
The two cases mentioned above produce a distribution of opinion clusters that conserve the initial mean opinion.
However, asymmetric distributions of closed-minded individuals are expected in 
Monte Carlo simulations because
of the finite number of individuals. These finite-size effects may 
easily break the symmetry around
the initial overall average opinion and lead to the formation of consensual 
states to the left or the right of 
$x=1/2$.

\subsubsection{Behavior in the presence of mass media ($m \ne 0$)}

\label{subsub:massmedia} 
\begin{figure}
\begin{center}
\mbox{\includegraphics[clip,angle=270,width=.6 \textwidth]{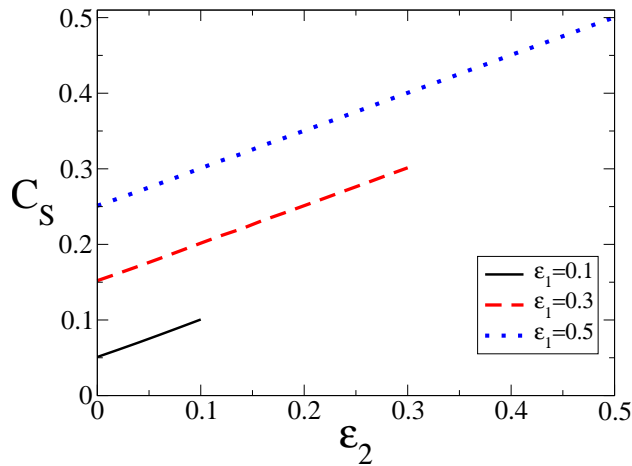}}
\end{center}
\caption{Normalised average size of the largest domain that includes all individuals whose opinion
is very close to the mass media, $C_{S}$, for $m=1.0$. This quantity is plotted as a function of $\epsilon_{2}$ 
for $\epsilon_{1}=0.1$ (Solid line), $0.3$ (Dashed line), and $0.5$ (Dotted line). Each data point is plotted
at $2 \times 10^{5}$ MCS after averaging over $100$ runs starting from uniformly distributed initial opinions
inside the interval $[0,1]$. The total number of individuals is $N=1000$.} \label{fig2new}
\end{figure}

In the presence of mass media, there is a competition between two social forces.
The tendency of the exogenous mass media to persuade individuals to follow it, versus the formation of 
spontaneous ordered states by endogenous social interactions. This problem has been already analysed
in systems of homogeneous individuals ($\epsilon_{i}=\epsilon$) and several 
interesting phenomena 
have been reported \cite{pineda, carletti, walter, avella, gargiulo}. However, as this 
work shows, the collective dynamic of the system becomes much more complex as 
we introduce heterogeneity, even in the present case where 
there are only two types of individuals,  $\epsilon_{1}\neq \epsilon_{2}$. 
For example, as we will show, the final ordered state strongly depends on the 
initial profile.

As mentioned in the introduction, we are particularly interested in analysing 
how significant is the effect of a mass media on our social system of 
heterogeneous individuals. Instead of calculating as usual the percentage or 
fraction of individuals whose opinion coincides with the opinion of the mass 
media, in this work we start from $100$ randomly distributed opinion 
profiles and once the system stabilizes, calculate, as a function of the 
parameters, how  often the external mass media persuades at least 
half of the population to follow its opinion, creating a majority. 
In this case, the so-called normalised average size of the largest domain that 
includes all the individuals whose opinion is very close to the 
mass media, $C_S$, coincides with the normalised average size of the largest 
domain defined previously (i.e. $C_{L}=C_{S}>0.5$).  

For $m=0$, there is not mass media acting on the system and the results of previous subsection are recovered 
(see Fig.~\ref{fig1}). For the opposite extreme case, $m=1$, the whole 
population only interacts with 
the media. Selecting $S=1$, the normalised largest domain increases with 
$\epsilon_{2}\in [0,\epsilon_{1}]$ as 
$C_L=C_{S}=(\epsilon_{1} + \epsilon_{2})/2$, for any 
$\epsilon_{1}\in [0,1]$ (see Fig.~\ref{fig2new}) . Thus, when $\epsilon_{1}=\epsilon_{2}=\epsilon$, the 
largest domain $C_{L}$ increases with the confidence level as expected, i.e. 
$C_{L}=C_{S}=\epsilon$ \cite{avella}. 

\begin{figure}
\begin{center}
\mbox{\includegraphics[clip,angle=270,width=.55 \textwidth]{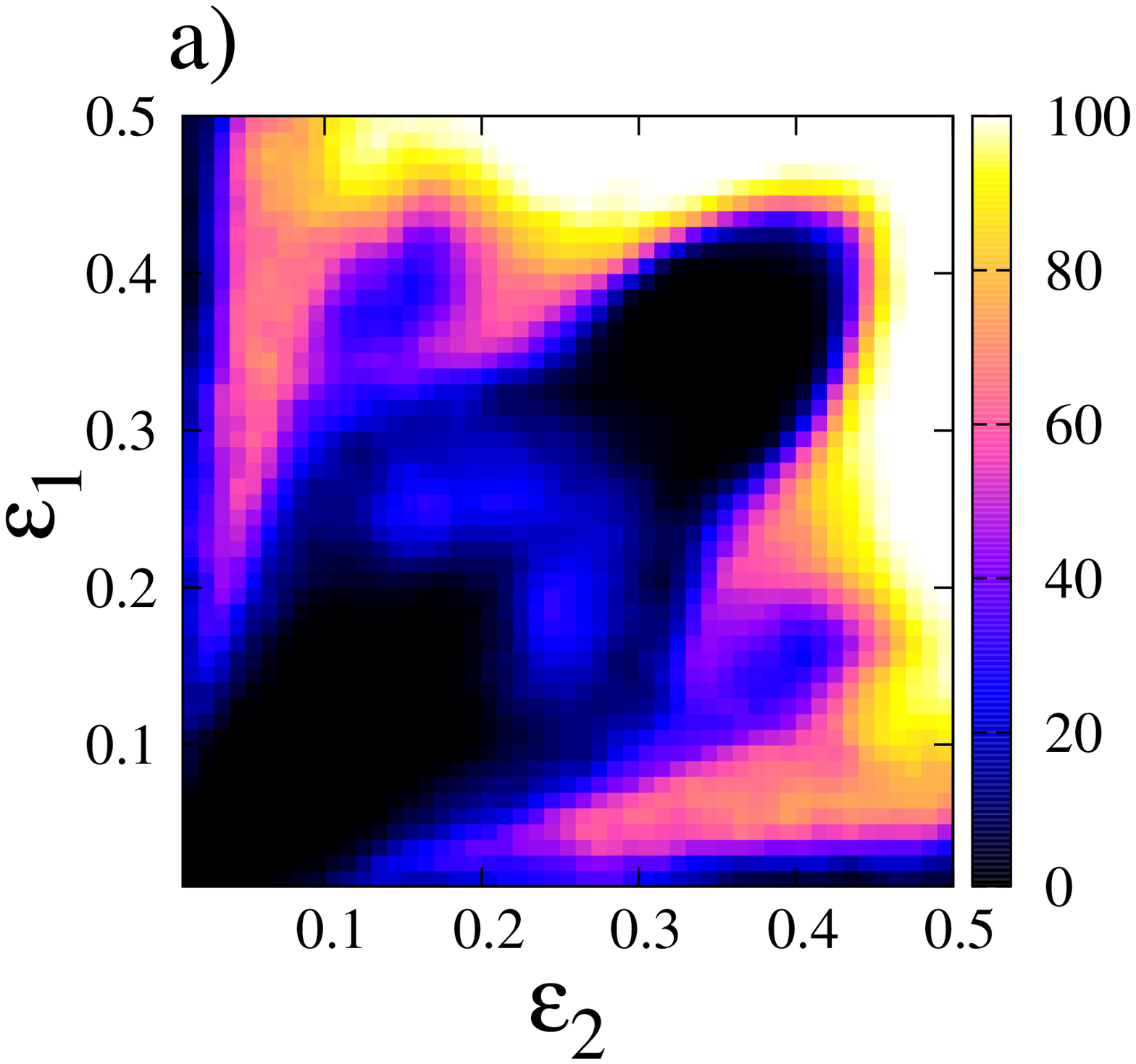}
\includegraphics[clip,angle=270,width=.55 \textwidth]{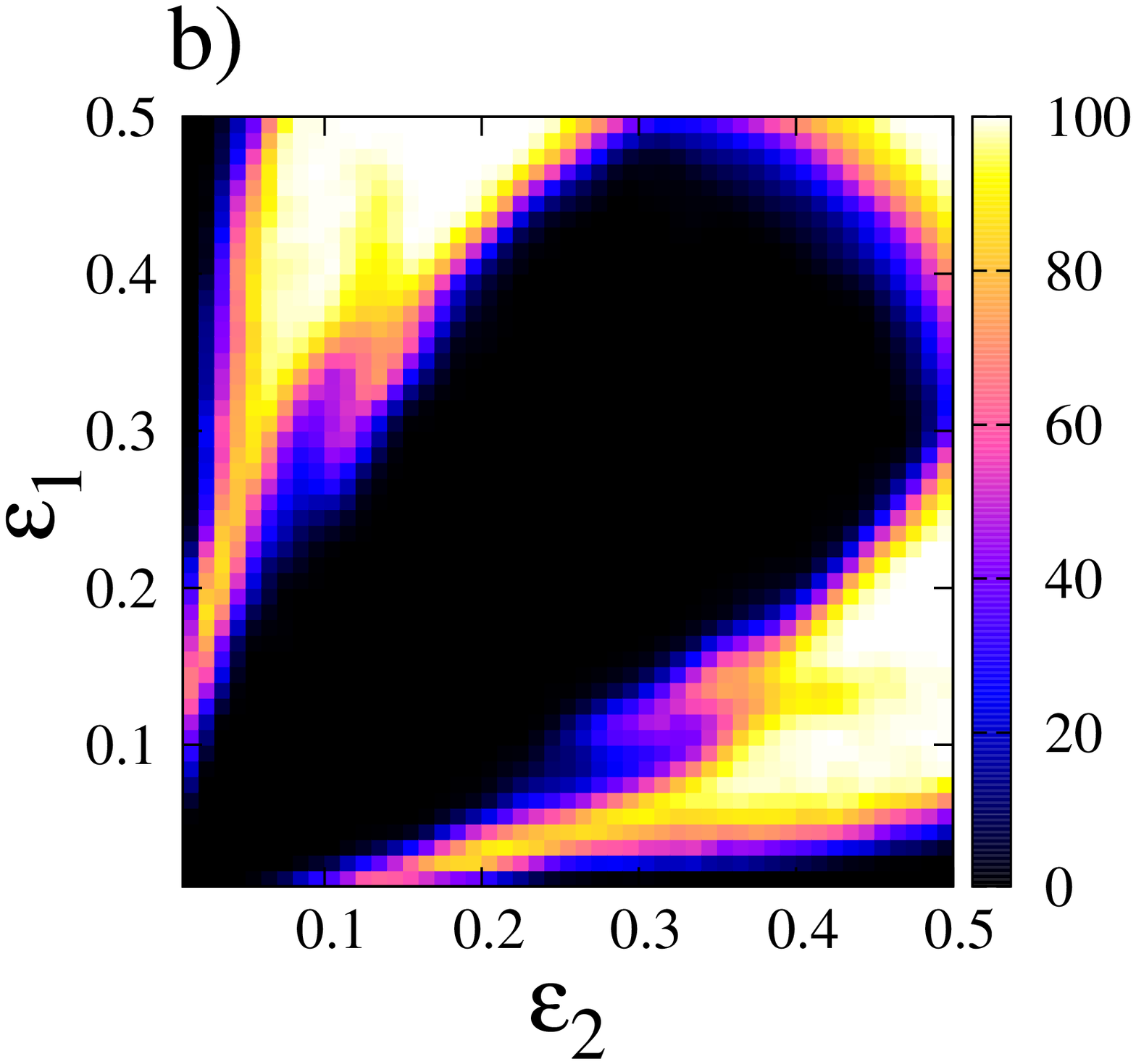}}
\caption{Heterogeneous DW model with two bounds of confidence ($\epsilon_{1}$ and $\epsilon_{2}$) and $m \neq 0$.
  Left panels show the case with $m=0.1$. Right panels show the case with $m=0.6$. a) and b) show the density plots
  of the frequency with which the mass media persuades more than half of the population to follow its opinion $S=1$. 
   In both figures the total number of individuals is $N=1000$ with each subsystem
 having $N/2$ individuals (the system is symmetric under the exchange of confidence levels). Each data point is plotted at
 $2 \times 10^{5}$ MCS after averaging over $100$ runs starting from uniformly distributed initial opinions inside the interval $[0,1]$. 
 }\label{fig3}
 \end{center}
\end{figure}

Figure~\ref{fig3} shows how efficient is the mass media to persuade more than 
half of the population to follow its opinion, in terms of $\epsilon_{1}$ and  
$\epsilon_{2}$.  The cases with $m=0.1$ and $0.6$ 
are shown in Fig.~\ref{fig3}a) and b), respectively. From these simulations 
it is clear that, in most of the cases and provided that the confidence levels 
are not too large, the mass media is unable to form a majority around its 
opinion when the system is too homogeneous (close to the diagonal line).
Similar to \cite{avella}, we found that parameter values from this region 
might lead to the formation of a consensus at an opinion different to that one of
the mass media. On the other hand, when one of the two
groups have a very small confidence level a highly disordered state is found 
(black 
regions close to the axes). Similar disordered states are found when the confidence values of the 
two groups are too small. However far from these two extreme cases, the 
persuasion capacity of the mass media improves. We can
conclude, that in our system the chance that the mass media has to
convince more that half of the population, is in general enhanced when the 
system is heterogeneous. In most the realisations, the mass media and the 
closed-minded individuals located
near $x=1$ act together and pull the rest of individuals toward that extreme 
position. From the parameter space diagrams, we can also
identify the counter-intuitive effect that a very strong mass media becomes 
inefficient to impose its    
opinion on the system. In our model this
phenomenon occurs when the two groups have large enough confidence values 
(see for example the region around  $\epsilon_{2}=0.45$ and $\epsilon_{1}=0.25$ 
in Figs.~\ref{fig3}a) and b)).
This effect, early reported in \cite{avella2}, is typical of systems with 
non-interacting states (or bounds). 
For large enough mass media and confidence levels,  the process of 
convergence to the mass media
message and the formation of an opinion cluster occurs faster;  this 
contributes to the early fragmentation of the
system and to limitate the number of individuals following the mass media 
$S=1$. However, for small values
of confidence levels,  the convergence process takes longer, allowing
enough local interactions to
take place and propagate the state of the mass media, leading to the growth of 
that state when $m$ increases.

\subsection{A society of agents with a wider distribution of bounded 
confidence levels}
\label{subsec:manybounds}
\begin{figure}
\begin{center}
\mbox{\includegraphics[clip,angle=270,width=.6 \textwidth]{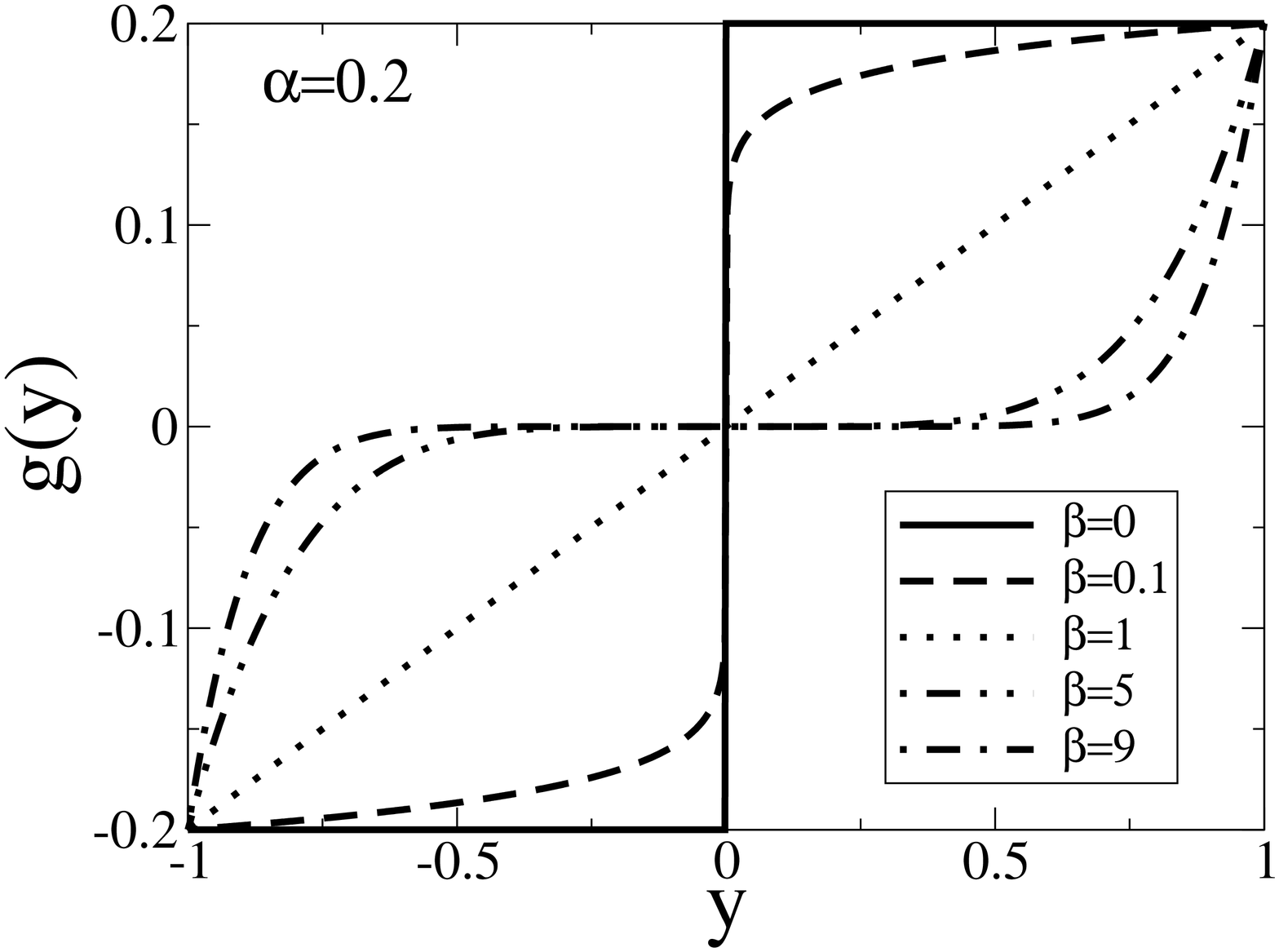}}
\end{center}
\caption{Function $g(y)=\alpha sign(y)\mid y \mid^{\beta}$, for $\alpha=0.2$, with $\beta=0$ (Solid line), 
$0.1$ (Dashed line),  $1$ (Dotted line), $5$ (dot-dot-dash line), and $6$ (dot-dash line). \cite{lorenz3}.}\label{fig4}
\end{figure}

Closer to reality is to consider situations were each individual 
of the society has her/his own particular confidence level. 
Following Lorenz \cite{lorenz3},  a general way to introduce different types of confidence levels is to assume that for each agent $i$
\begin{equation}
 \epsilon_{i}=\epsilon_{i0}+g(y_{i}),
 \label{confidence}
\end{equation}
where
\begin{equation}
 g(y_{i})=\alpha sign(y_{i})\mid y_{i} \mid^{\beta},
 \label{function}
\end{equation}
with $y_{i}$ distributed between $-1$ and $1$ \cite{lorenz1}. For simplicity in this 
work we will assume that $\epsilon_{i0}=\epsilon_{0}$. The parameter 
$\alpha$, which runs form $0$ to $\epsilon_{0}$, represents the range of 
heterogeneity. The parameter $\beta$ characterises the width of the 
distribution and in this work will run from zero to 
$9.9$. A general form of Eq.~\ref{function} is plotted in Fig.~\ref{fig4}, for a fixed 
value of $\alpha$ and several values of $\beta$. When $\beta=0$, individuals can only have 
two possible confidence levels, either $\epsilon_{0}-\alpha$ or $\epsilon_{0}+\alpha$. In this case, we recover
results presented in subsection~\ref{subsec:twogroups} \cite{lorenz2}. When 
$\beta>0$, the confidence values are distributed along the interval 
$[\epsilon_{0}-\alpha,\epsilon_{0}+\alpha]$, with large values of $\beta$ 
favoring intermediate confidence levels, $\epsilon_{0}$, (see dot-dash line in 
Fig.~\ref{fig4}).  

\subsubsection{Behavior in the absence of mass media ($m=0$)}
\label{subsubsec:nomassmany}

\begin{figure}
\begin{center}
\mbox{\includegraphics[clip,angle=270,width=.45 \textwidth]{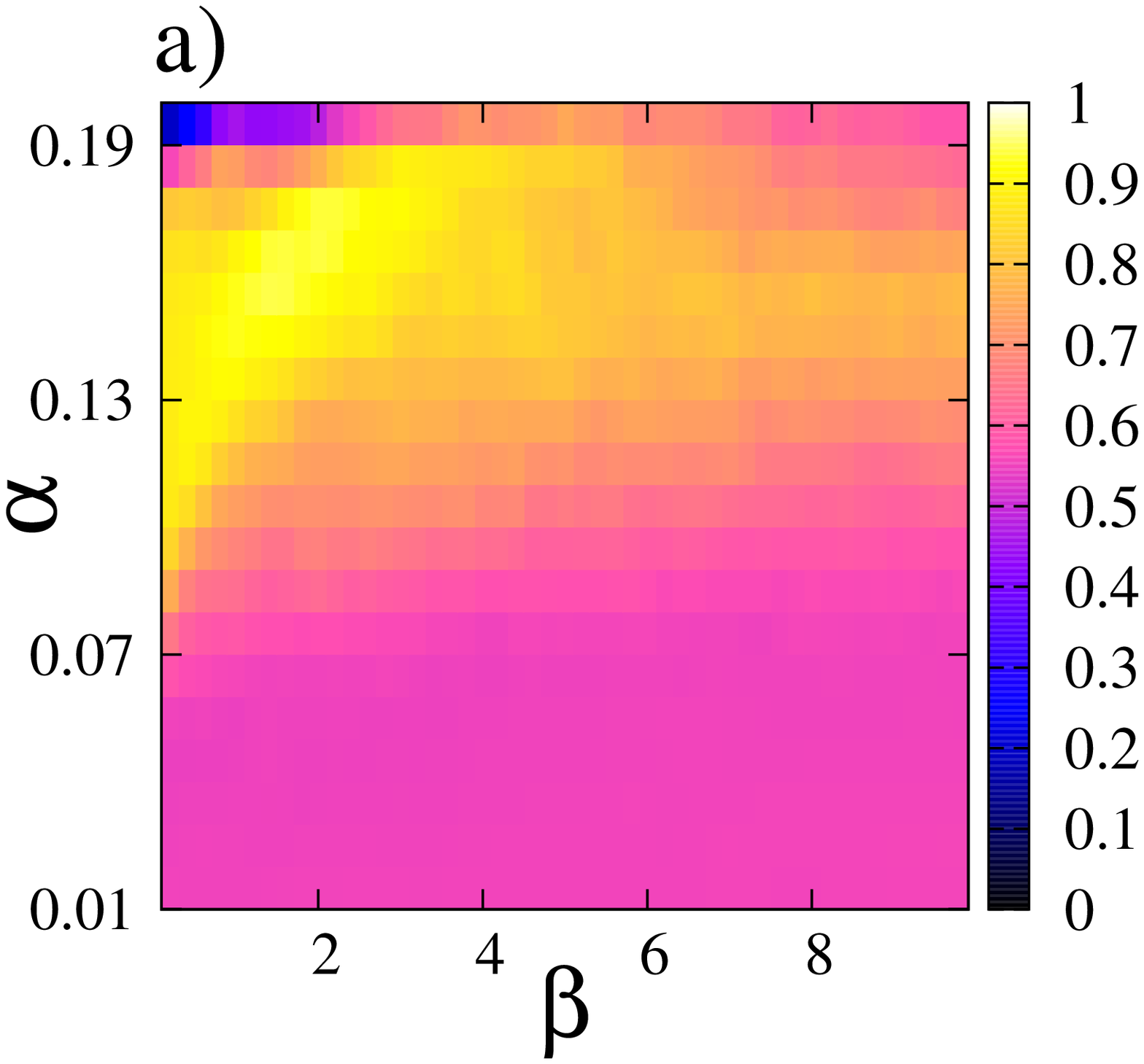}}
\hspace{5mm}
\mbox{\includegraphics[clip,angle=270,width=.45 \textwidth]{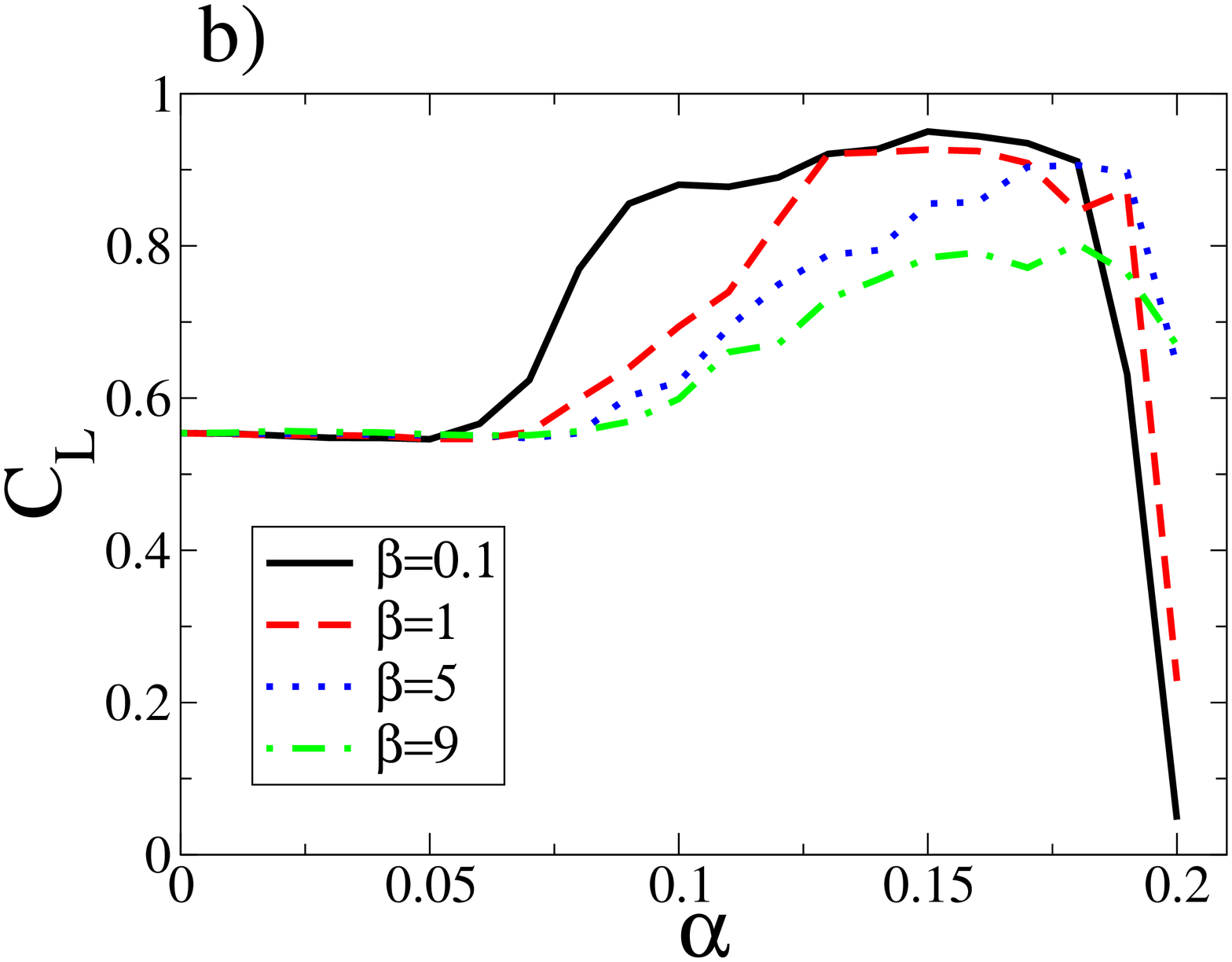}}
\end{center} 
\caption{Heterogeneous DW model without mass media.  a) Density plot of the 
normalised average size of the largest domain, $C_L$, in the plane ($\beta$, 
$\alpha$). b) The same parameter, $C_L$, as a function of $\alpha$ for $\beta=0.1$ (Black solid line), $1.0$ (Red dashed line),
$5$ (Blue dotted line), and $9$ (green dot-dash line). The total population is $N=1000$. Each data point is plotted at
 $2 \times 10^{5}$ MCS after averaging over $100$ runs starting from uniformly distributed initial opinions inside the interval $[0,1]$.
 In all cases we assume $\epsilon_{0}=0.2$.}
\label{fig5}
\end{figure}

In this subsection, we are interested in the behavior of the normalised average 
size of the largest domain $C_L$ of the system, as a function of the two new 
control parameters. In Figure~\ref{fig5}a) we plot the magnitude of $C_L$ on 
the plane ($\beta, \alpha$), for $\epsilon_{0}=0.2$.  When $\beta=0$, there are not 
intermediate confidence levels. We get  a system composed by two 
groups characterised by $\epsilon_{0}=0.2 \pm \alpha$. The parameter $\alpha$ 
determines how different are the confidence levels of the two groups. For 
example, when $\alpha=0.2$, we end up with a group of opened-minded individuals 
($\epsilon=0.4$) and a group of closed-minded individuals ($\epsilon=0$), of 
approximately the same size. This combination gives raise to a 
disordered state similar to the one observed in Fig.~\ref{fig1} (Top left corner in Fig.~\ref{fig5}).  
When $\alpha=0$, the system is homogeneous with $\epsilon_{i}=\epsilon_{0}=0.2$. For this value of 
$\epsilon$, the typical pattern is characterised by two almost identical opinion clusters ($C_{L} \approx 0.5$) \cite{lorenz1} .
For other values of $\alpha$, one can verify that the behaviour is similar to the one 
obtained in Sec. $1$. In particular, a state of total consensus 
toward an extreme is identified for
 intermediate values of $\alpha$. Which extreme is chosen by the majority 
depends on the initial profile of 
 opinions.

For large enough values of $\beta$, the function $g(y)$ is most of the time very close to zero, 
and the number of individuals with a confidence level, $\epsilon_{0}=0.2$, increases. As mentioned before, this situation
gives raise to a state of polarization, where $C_{L} \approx 0.5$. This 
polarized state is 
characterised by two almost identical opinion clusters  symmetrically placed 
around the center of the opinion space or with
a small shift towards one of the extreme positions. However, the most interesting finding is that there still exist intermediate values
 of $\beta$ for which the chance to reach a complete consensus 
toward the extremes can be considerably improved, provided  $\alpha$ acquires moderate values. If $\alpha$ is 
too small we get a polarized state  $C_{L} \approx 0.5$. On the other hand, if 
the heterogeneity is too large, the system can go back to a state of 
polarization 
or to a disordered state depending of $\beta$. Intermediate values 
of $\alpha$ assure the 
presence of individuals with levels of confidence necessary to induce 
consensual
states toward the extremes. Figure~\ref{fig5}b) shows $C_{L}$ as a function of 
$\alpha$ for several values of 
$\beta$. Note that, as expected, for $\alpha=0.2$ the value of $C_{L}$ increases from zero (disorder)
to approximately $0.5$ (polarization) as $\beta$ increases. It also shows that 
for 
intermediate values of $\alpha$ the chance to reach a consensus is improved. If 
$\alpha$ decreases below a threshold value, a polarized state $C_{L} \approx 
0.5$ is recovered again.

\subsubsection{Behavior in the presence of mass media ($m \ne 0$)}   
\label{subsubsec:massmany}
\begin{figure}
\begin{center}
\mbox{\includegraphics[clip,angle=270,width=.45 \textwidth]{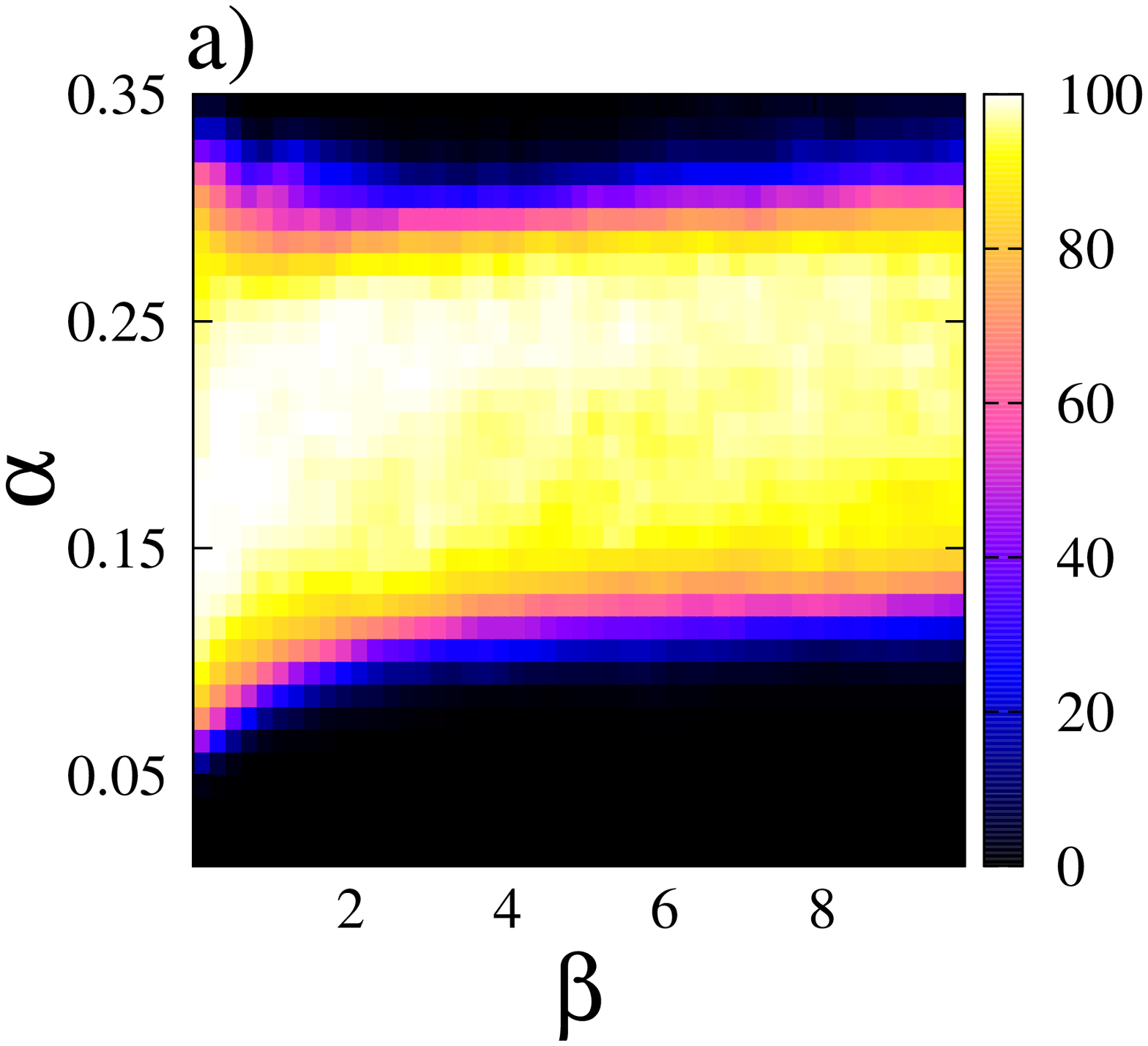}}
\hspace{5mm}
\mbox{\includegraphics[clip,angle=270,width=.45 \textwidth]{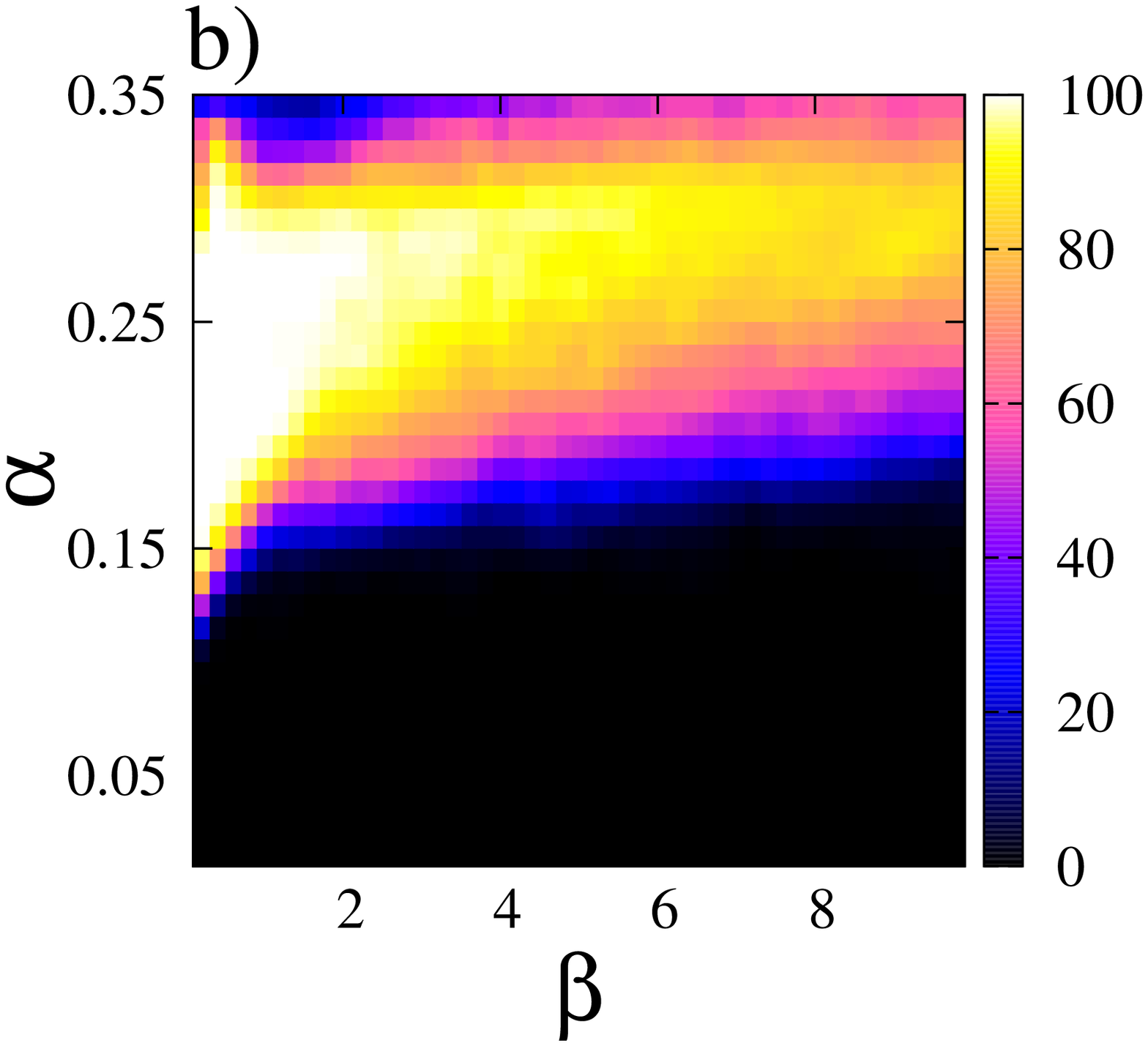}}
\end{center}
\caption{Heterogeneous DW model with mass media. Density plots, in the plane 
($\beta$, $\alpha$), of the frequency with which the mass media persuades more 
than half of the population to follow its opinion. a) $m=0.1$ . b) $m=0.6$. In 
both figures the total number of individuals is $N=1000$ and $\epsilon_{0}=0.35$.
Each data point is plotted at $2 \times 10^{5}$ MCS after averaging over $100$ runs
starting from uniformly distributed initial opinions inside the interval $[0,1]$.}
\label{fig6}
\end{figure}

In this case, we calculate the frequency with which the mass media persuades more 
than half of the population to follow its opinion for two values 
of $m$. After long enough simulation times, 
only steady states characterised by $C_{L}>0.5$ are counted.  
Figures.~\ref{fig6}a) and \ref{fig6}b) show this persuasion frequency for a 
lower value of $m$, $m=0.1$ and a larger one, $m=0.6$, respectively. 
We choose $\epsilon_{0}=0.35$ to assure that in the absence of 
heterogeneity ($\alpha=0$) we are in a region where the mass media does not 
have any appreciable effects (see Fig.~\ref{fig3}), since we want to quantify the impact of this 
exogenous factor when increasing heterogeneity.  For $m=0.1$, there is a well 
defined region for intermediate values of the heterogeneity parameter $\alpha$ 
where  the persuasion of the mass media is optimal, meaning a high chance of 
reaching consensus around $S=1$. Similar to Fig.~\ref{fig3}, the mass media acts in 
accordance with some closed-minded individuals and pull the rest of individuals 
towards their extreme view, 
forming a consensual state of extremists.
For higher $\alpha$ values, the heterogeneity 
becomes too large and the mass media can not impose its state.
For $m=0.6$, although the intensity of the mass media is higher that the 
previous one, its capacity of persuasion
 is reduced, indicating that, in order to have optimal persuasion for a large range of parameters, the mass media intensity must be 
not too large. As mentioned in previous sections, this phenomenon 
occurs because $m$ and $\epsilon_{o}$ are too big, leading to
a rapid fragmentation of the system.

\section{Comparison with the heterogeneous Hegselmann-Krause (HK) model with mass media}
\label{sec:HK}

\begin{figure}
\begin{center}
\mbox{\includegraphics[clip,angle=270,width=.5 \textwidth]{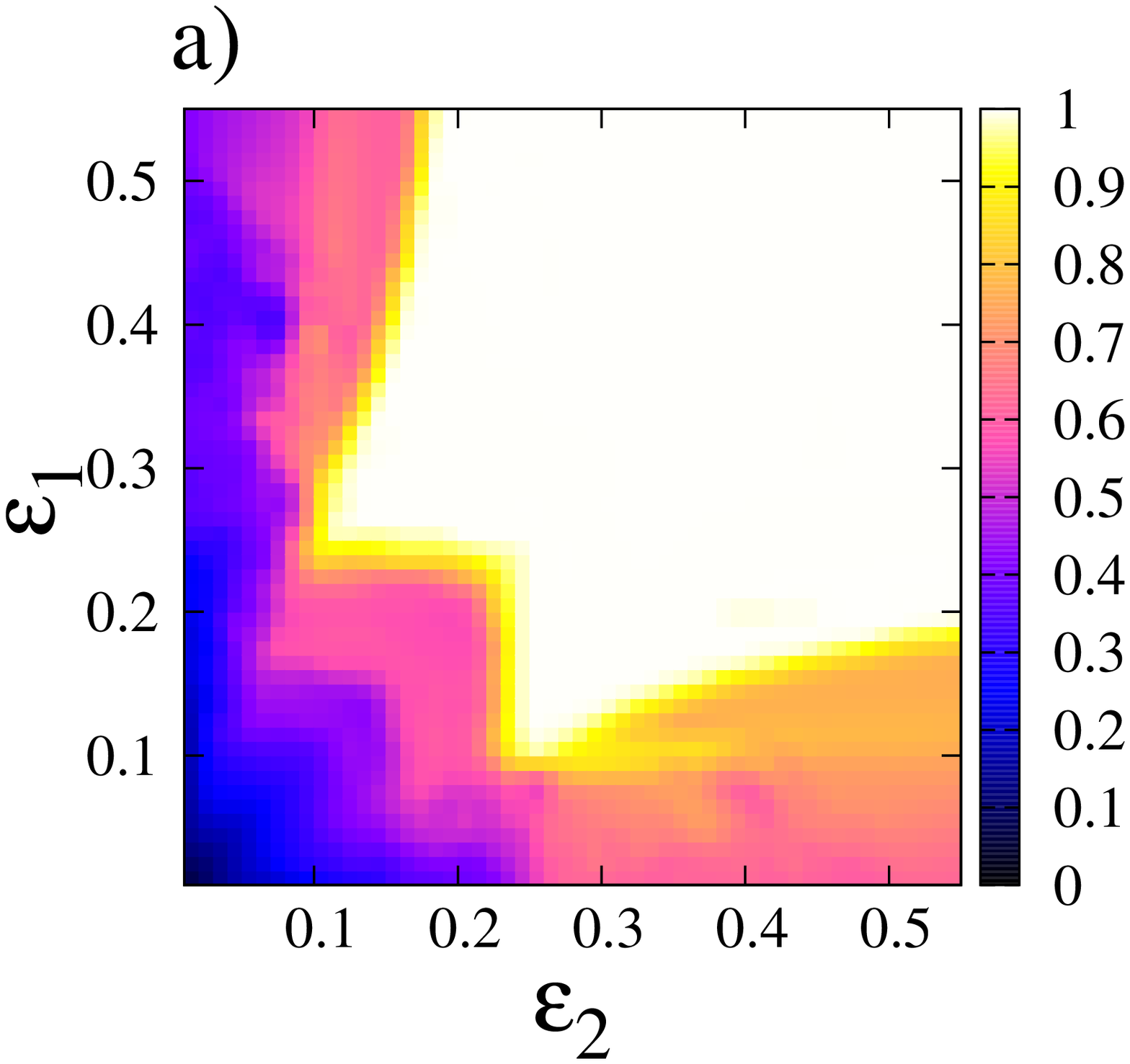}
\hspace{2mm}
\includegraphics[clip,angle=270,width=.5 \textwidth]{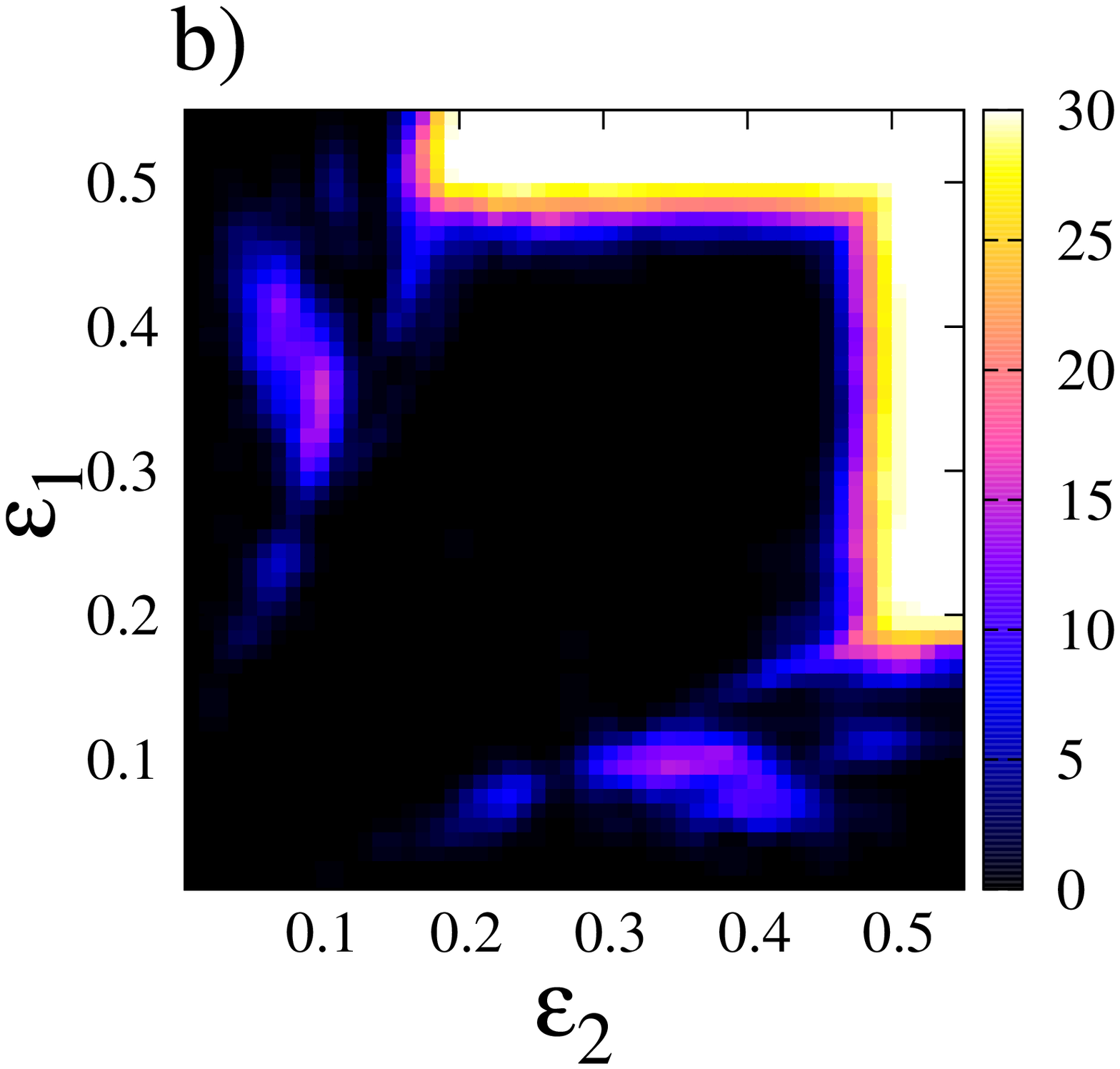}}
\hspace{2mm}
\mbox{\includegraphics[clip,angle=270,width=.5\textwidth]{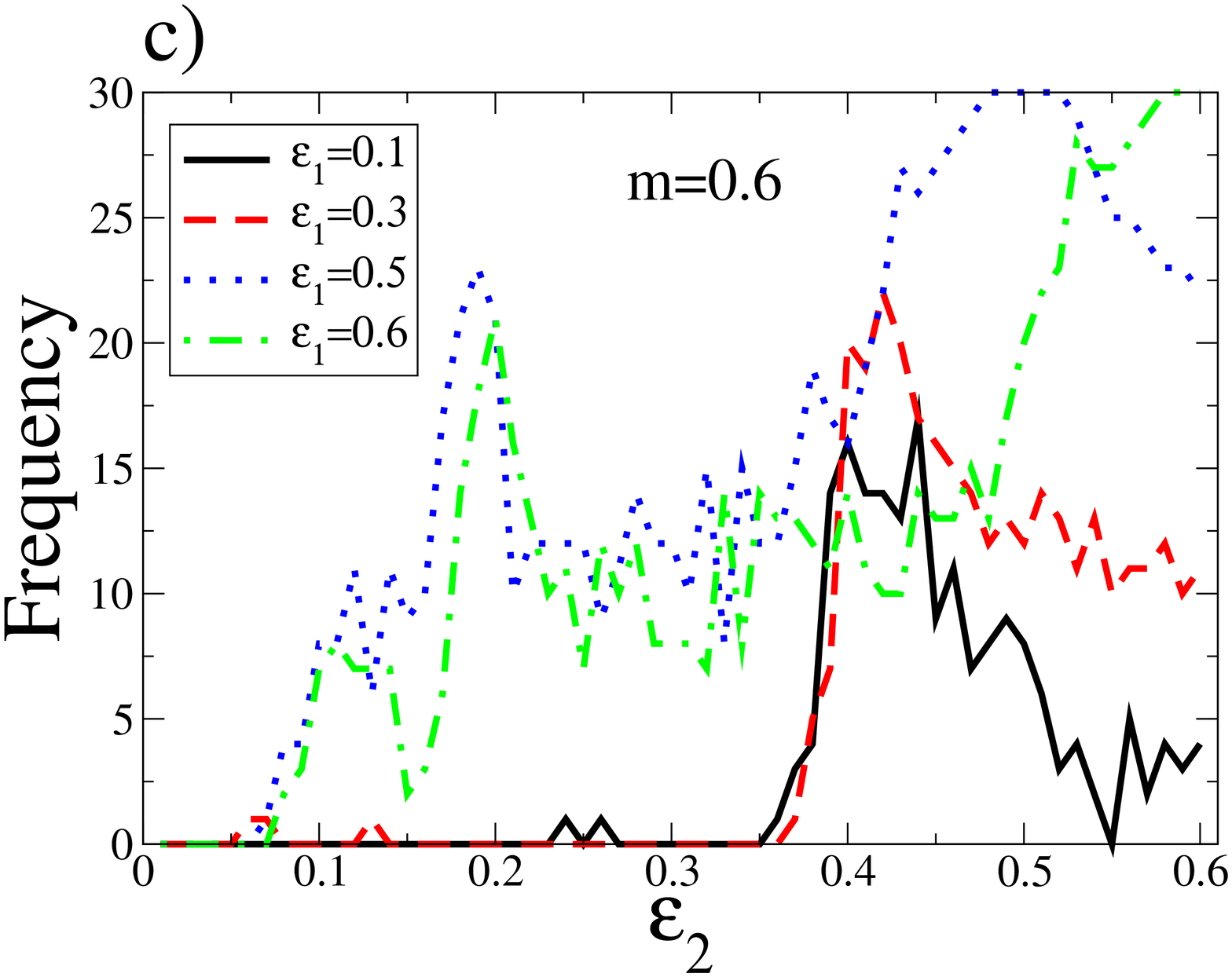}}
\end{center} 
\caption{Heterogeneous HK model. a) Density plot of the 
normalised average size of the largest domain, $C_L \in [0,1] $, in the plane 
($\epsilon_{2}$, $\epsilon_{1}$) for $m=0$. b) Density plot of the 
frequency with which the mass media persuade more than half of the
population to follow its opinion in the plane ($\epsilon_{2}$, $\epsilon_{1}$) for
$m=0.1$. c) Frequency with which the mass media persuade more 
than half of the population to follow its opinion, for $m=0.6$. Here, this quantity is plotted
as a function of $\epsilon_{2}$ for $\epsilon_{1}=0.1$ (Black solid line),  $\epsilon_{1}=0.3$ (Red dashed line),
$\epsilon_{1}=0.5$ (Blue dotted line), and  $\epsilon_{1}=0.6$ (Green dot-dash line). 
 In both figures the total number of individuals is $N=1000$ with each 
subsystem having $N/2$ individuals.The behaviour is  symmetric under the exchange
of confidence levels. Each data point is plotted at $7 \times 10^{3}$ MCS after averaging over $30$ runs
starting from uniformly distributed initial opinions inside the interval $[0,1]$. }\label{fig8}
\end{figure}

\begin{figure}
\begin{center}
\mbox{\includegraphics[clip,angle=270,width=.5 \textwidth]{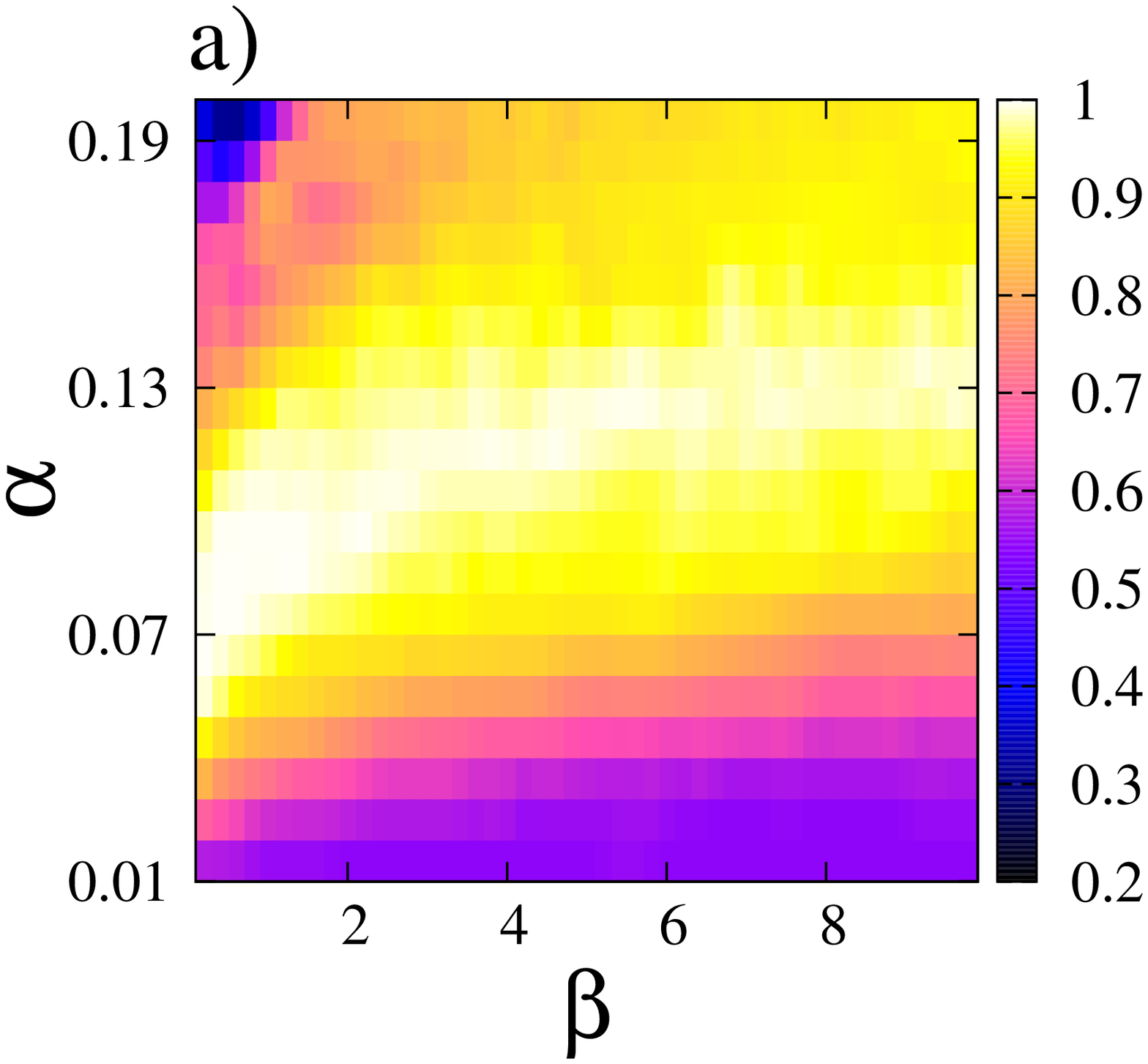}
\includegraphics[clip,angle=270,width=.5 \textwidth]{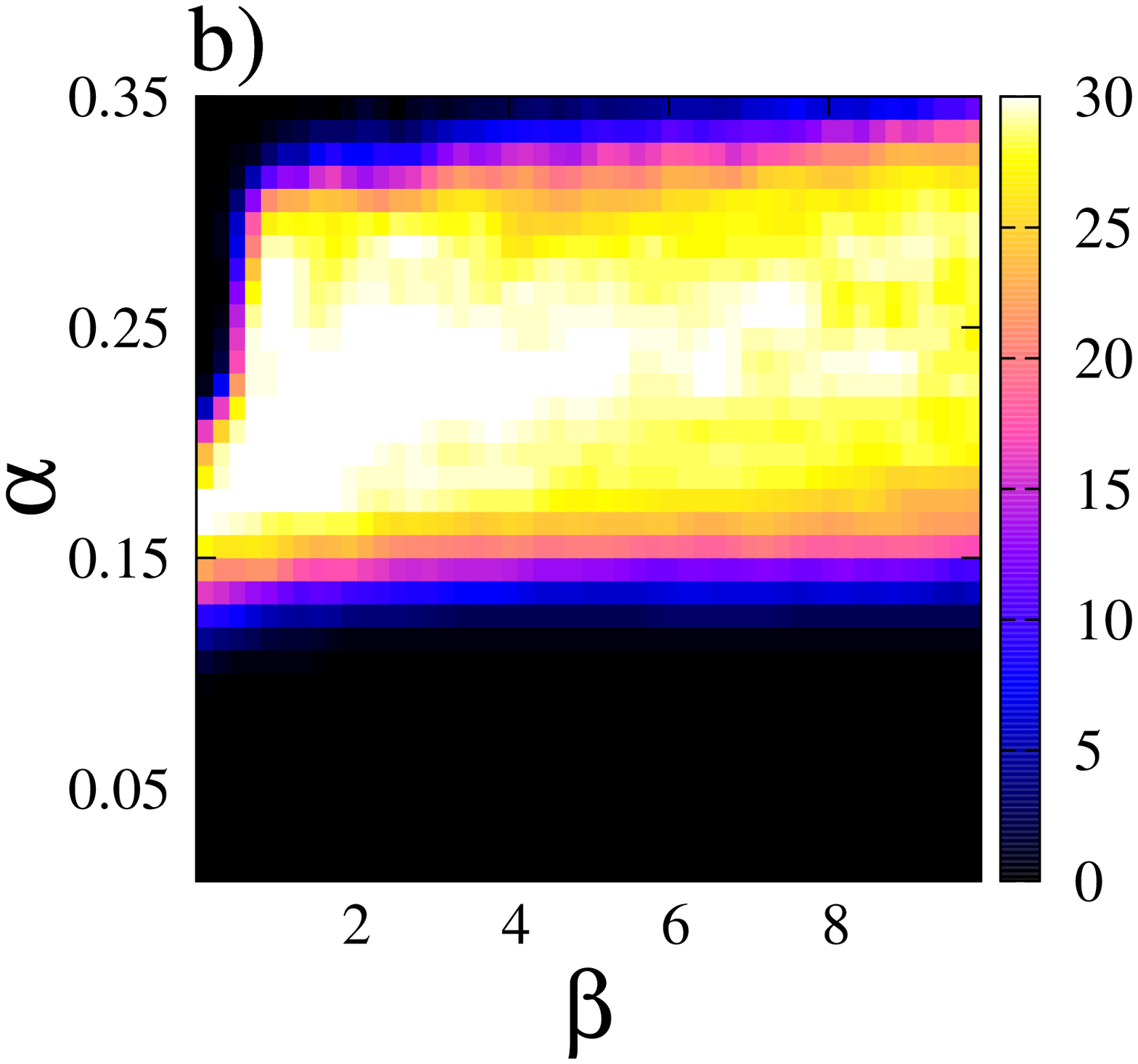}}
\mbox{\includegraphics[clip,angle=270,width=.5 \textwidth]{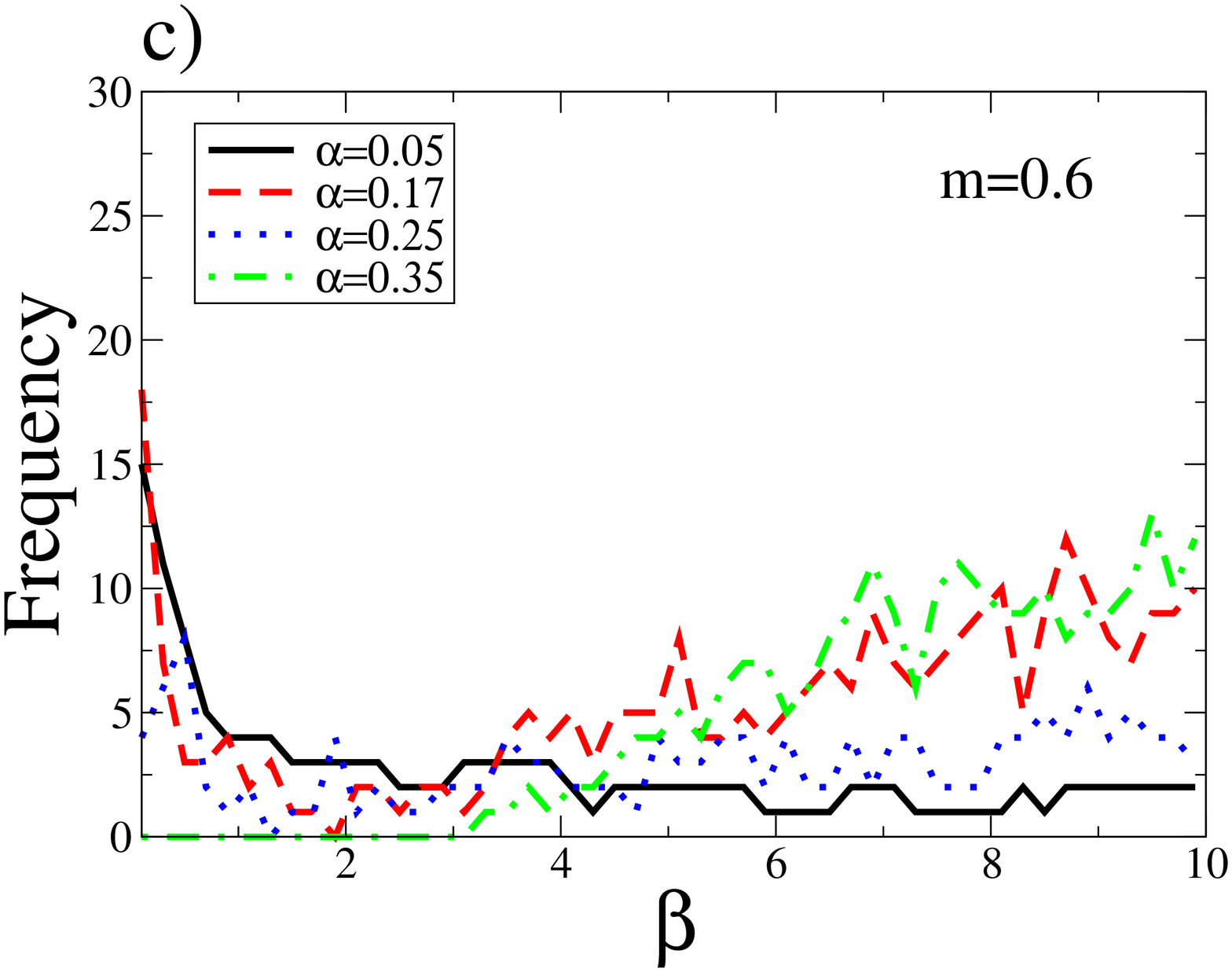}
}
\end{center} 
\caption{Heterogeneous HK model. a) Density plot of the 
normalised average size of the largest domain, $C_L \in [0,1] $,  in the plane ($\beta$, $\alpha$), for
$m=0$ b) Density plot of the  frequency with which the mass media persuade more 
than half of the population to follow its opinion in the plane ($\beta$, $\alpha$), for
$m=0.1$. c) Frequency with which the mass media persuade more 
than half of the population to follow its opinion, for $m=0.6$. Here, this quantity
is plotted  as a function of $\alpha$ for $\alpha=0.05$ (Black solid line),  $\alpha=0.17$ (Red dashed line),
 $\alpha=0.25$ (Blue dotted line), and  $\alpha=0.35$ (Green dot-dash line). 
 In both figures the total number of individuals is $N=1000$, with each 
subsystem having $N/2$ individuals.
Each data point is plotted at $7 \times 10^{3}$ MCS after averaging over $30$ runs
starting from uniformly distributed initial opinions inside the interval $[0,1]$. }\label{fig9}
\end{figure}

The HK model differs from the DW model in the fact that the relation 
between individuals is not based into a one by one interaction. Instead, in 
the HK model at each time step an 
individual interacts with all the individuals whose opinions lie 
into its area of confidence \cite{lorenz1}. After this interaction the 
individual takes the 
average opinion of the group. In certain way this model can describe the 
so-called herd mentality. Starting with a random distribution in the opinion
space $ x^{i}_{0} \in [$0$,$1$]$, at time $n$ an individual $i$ is selected at random:

1) With probability $m$, if $|x^{i}_{n}-S| < \epsilon_{i}$, $i$ interacts with the field $S$ such that,  
\begin{equation}
x^{i}_{n+1}=\frac{x^{i}_{n}+S}{2};
\end{equation}
if $|x^{i}_{n}-S| \ge \epsilon_{i}$ the opinion of the individual does not change. In any case the cycle starts again.

2) With probability $1-m$ the individual takes the average opinion of the group of individuals inside the confidence range, such that 
\begin{equation}
x^{n+1}_{i}=\frac{1}{N_i}\sum\limits_{j:|x^{n}_{i}-x^{n}_{j}| < 
\epsilon_{i}}x^{n}_{j},
\end{equation}
where $N _i$ is the number of individuals whose opinion lie into 
the field of confidence of $i$. That means all the $j$ individuals that satisfy 
$|x^{n}_{i}-x^{n}_{j}| < \epsilon_{i}$.  The sum includes $i$. In simple 
words, individual $i$ takes the average opinion of the individuals that think 
like him. In any case the cycle starts again.

The procedure is repeated by selecting at random another individual, and so on. 
As before, we consider that the iteration time increases as 
$n \rightarrow n+1$, and introduce a time variable $t=n\Delta 
t$,  $\Delta t=1/N$,  where the time unit is  given in terms of Monte Carlo 
steps 
($MCS$).  In our simulations, we choose as mass media "opinion", $S=1$. The 
probability $m \in [0,1]$  represents, as before, the strength of 
the mass media compared with the personal interactions.  Note that, contrary to 
the case of the DW model, this model does not have any convergence parameter. 
In the original version of the HK model the update was done synchronously (all opinions
are updated each iteration time $n$) \cite{krause0}. As described, in our simulation the individuals are chosen 
randomly. This change may affect the convergence time to the final steady state \cite{pineda0}.

In this section, the results are averages over $30$ runs starting from randomly distributed initial opinions. Then the frequency, defined as the number of runs for which the mass media persuades more than half the population to follow its opinion, has a maximum value of $30$.
In the case of only two bound of confidences, $\epsilon_{1}$ and 
$\epsilon_{2}$, and no mass media ($m=0$),  the HK model also shows the 
remarkable phenomenon of consensus far below the consensus threshold of the homogeneous 
case (see Fig.~\ref{fig8}a)), that for the system size we are considering here
its roughly around $\epsilon_{1}=\epsilon_{2}=\epsilon \approx 0.25$. The main 
difference with the DW model is that the parameter space region where this phenomenon
occurs is reduced. 

When $m>0$,  the persuasion capacity of the mass media also depends very much on the parameters values.
For some set of parameters, the mass media is able to persuade more than half of the
population to follow its opinion, in almost all realisation. But, there are also regions on the parameter
space where this persuasion in not optimal. Here, we are interested in quantify the number of times the mass
media convinces more than half of the population to follow its opinion ($C_{S}>0.5$). As an example, Fig.~\ref{fig8}b)
shows that, when $m=0.1$, there is a well defined region on the plane ($\epsilon_{2}$, $\epsilon_{1}$) inside
which the persuasion capacity is maximal. However, Fig.~\ref{fig8}c) shows that, if $m=0.6$, the mass media
rarely convinces more than half of the population.  

Figure~\ref{fig9}a) shows the case of a society of agents with a heterogeneous 
distribution of bounded confidence levels, for the case $m=0$ \cite{lorenz3} . Here, $\epsilon_{i}$ is given 
by Eq.~4 (see subsection~\ref{subsec:manybounds}). As expected, the capacity of the HK 
model to reach consensus for  intermediate values of heterogeneity is higher 
than in the DW model. In the HK model the interaction between 
individuals is global rather than local. For $m=0.1$ (see Fig.~\ref{fig9}b)), 
we also find a region for  intermediate values of heterogeneity where the 
persuasion capacity of the mass media is optimal. However, Fig.~\ref{fig9}c)
shows that the persuasion capacity is strongly affected when $m=0.6$. The number 
of times the mass media persuades more than half of the population does not
overcome the barrier of $20$ (from a maximum of $30$), for any $\alpha$ value considered.  
The parameters $m$ and $\epsilon_{o}$ are too big, leading to a 
rapid fragmentation of the system, and to an early separation of the cluster with 
state $S=1$ from the other clusters.

\section{Summary and conclusions}
\label{sec:conclusion}
In this work we have studied the effect of introducing an external message in
the Deffuant et al. and Hegselmann and Krause models for continuous opinion 
dynamics with heterogeneous bounds of confidence.  We started by analysing a 
system of agents composed by two groups differing only in their confidence 
levels. Then, we studied a case  where each agent has her/his own characteristic level of 
confidence.

In the case of the Deffuant et al. model with two bounds of confidence, 
$\epsilon_{1}$, $\epsilon_{2}$ and no mass media, $m=0$, we found that the 
chances to reach consensus are improved far from the diagonal of symmetry 
($\epsilon_{1}$= $\epsilon_{2}$) where the homogeneous version of the Deffuant 
et al. model is recovered. This consensus state exists even for values of 
confidence levels far below the consensus threshold of the opposite homogeneous situation \cite{lorenz2}.  
The new consensus state is also located at extreme locations in 
the opinion space. We performed a master equation analysis of the system and concluded that in order to get such a consensual state, 
a group of closed-minded individuals located near the extremes are needed. One can also inference that the origin of the 
increase of the largest cluster size when the tolerance parameter is distributed rather than constant, is due to a subtle interplay
between closed- and open-minded individuals. Open-minded individuals can pull closed-minded individuals towards other opinion
clusters of closed-individuals and vice versa.

To analyse what occurs when the mass media is present, we calculated, for one hundred homogeneous 
and random initial conditions, how often the mass media persuades more than half of the population to follow its opinion. 
We found that this quantity is very sensitive to the initial conditions and control parameters. 
In particular, this quantity reveals that the response of the system to the external message is
optimal  when the confidence levels of the two groups are sufficiently  different. 
Our simulations show that, if they are too similar, the external message does not have a strong capacity to 
convince more than half of the population (to have an idea of what occurs in 
the diagonal line or homogeneous case we refer to \cite{avella}). On the other 
hand, if one of the groups has a very small confidence level, the other group 
composed by open-minded agents, will always be interacting with a noisy 
environment created by agents that never change their opinions. This leads to 
a state of disorder or pluralism. When a society with a heterogeneous distribution
of bounded confidence levels but without the influence of any mass media is 
considered, we also found that intermediate values of heterogeneity enhance the 
chance to reach consensus. Similarly, the chance of the mass media to persuade more than 
half of the population to adopt its message is only improved 
provided we consider intermediate values of heterogeneity.
The counter-intuitive effect, in which if the mass media intensity 
is too large, it becomes unable to impose its message to the system, is also presented
in this version of the DW model. This finding supports early studies using 
Axelrod's model of dissemination of culture \cite{avella2}.

With respect to the Hegselmann and Krause model we found that diversity in 
bounds of confidence also enhances the chance to reach consensus. In the case
 of two bounds of confidence and no mass media, states of consensus are found
 below the critical consensus threshold of the homogeneous case. 
 When the system is under the influence of the mass media, we tested how
 often the mass media persuades more than half of the population to follow its opinion.
 We performed simulations with just thirty different random initial conditions, and verified
 how many of these initial conditions end up with more than half of the population following
 the mass media opinion. The main finding was that, as in the DW model, the system is very sensitive
 to the initial profile and parameter values. When a heterogeneous distribution of bounded confidence 
levels where considered, the final behaviour was very similar to that of the DW model. Consensus and optimal capacity
of persuasions were found for intermediate values of heterogeneity and mass media intensity.

Both models present similar phenomena when the mass media intensity, $m$, is null, small, or medium. 
However, the main difference between them is clearly appreciated when large enough mass media intensities are considered.
In particular, the persuasion capacity of the mass media is lower in the Hegselmann-Krause model. 
In contrast to the Deffuant et al. model where all interactions between individuals are on one-to-one basis,
 in the Hegselmann-Krause model considered in this work, individual interactions are global.  
 This global communication pathway is an important ingredient to effectively resist the impact of the mass media,
 particular for high intensities.
  
In this article we stress the importance that the interplay between 
heterogeneous bounds of confidence and mass media has in continuous opinion dynamic. 
Extension of this work should consider complex networks \cite{avella,vazquez}. We also notice that one can consider different forms of introducing
heterogeneity into the models. It may affect in different
ways the chance to reach consensus and the optimal response to external stimulus in continuous opinion dynamics.
The role of noise and initial conditions is also an important issue to address \cite{pineda0,toral1}.
It may be interesting to analyse in detail what happens in the 
regions of the parameters where the mass media
does not convince the majority of the population, as they may be related to the 
so-called states different to that of the mass media \cite{avella}. It would also
be interesting to expand the analysis by 
introducing other rules such as a global interaction with the
 external mass media or to introduce different communication rules between 
individuals. This would allow the model to carry 
out studies of more realistic opinion dynamic behaviours \cite{maxi}.   
\section{Acknowledgments}
M. P acknowledges support by USB-DID through the project S1-IN-CB-010-12.
 \appendix

\section{Master equation for DW with two bounds of confidence and $m=0$}
\label{me}

In this section, we study by a mean-field approximation a society of 
individuals with
two bounds of confidence, in the absence of mass media.
Let $P_{1}(x,t)$ and  $P_{2}(x,t)$ be the probability density functions of 
individual opinions 
in a system with bounds of confidence $\epsilon_{1}$ and 
 $\epsilon_{2}$, at time $t$, respectively. In analogy with classical kinetic 
theory, the time evolution of these probabilities obeys a
set of coupled master equations, given by

\begin{equation}
  \frac{\partial P_{1}(x,t)}{\partial t}=P_{11}(x,t)+P_{12}(x,t),
 \label{eq:et}
\end{equation}
\begin{equation}
  \frac{\partial P_{2}(x,t)}{\partial t}=P_{22}(x,t)+P_{21}(x,t),
 \label{eq:eqi}
\end{equation}
with
\begin{eqnarray} 
P_{11}(x,t) & = & 4\int_{|x-x'|<\epsilon_{1}/2}dx'P_{1}(2x-x',t)P_{1}(x',t) \nonumber \\
  & & - 2P_{1}(x,t)\int_{|x-x'|<\epsilon_{1}}dx'P_{1}(x',t)],
 \label{eq:defuantdiff11}
\end{eqnarray}

\begin{eqnarray}
P_{12}(x,t) & = & 2\int_{|x-x'|<\epsilon_{1}/2}dx'P_{1}(2x-x',t)P_{2}(x',t) \nonumber \\
  & & - P_{1}(x,t)\int_{|x-x'|<\epsilon_{1}}dx'P_{2}(x',t)],
 \label{eq:defuantdiff12}
\end{eqnarray}
and
\begin{eqnarray}
P_{22}(x,t) & = & 4\int_{|x-x'|<\epsilon_{2}/2}dx'P_{2}(2x-x',t)P_{2}(x',t) \nonumber \\
  & & - 2P_{2}(x,t)\int_{|x-x'|<\epsilon_{2}}dx'P_{2}(x',t)],
 \label{eq:defuantdiff22}
\end{eqnarray}
\begin{eqnarray}
P_{21}(x,t) & = & 2\int_{|x-x'|<\epsilon_{2}/2}dx'P_{2}(2x-x',t)P_{1}(x',t) \nonumber \\
  & & - P_{2}(x,t)\int_{|x-x'|<\epsilon_{2}}dx'P_{1}(x',t)],
 \label{eq:defuantdiff21}
\end{eqnarray}
The integrals over $x'$ run over the interval $[0,1]$. An important property of these equations 
is that if the initial conditions are symmetric around the central point 
$x=1/2$, namely that $P_{1}(x,t=0)=P_{1}(1-x,t=0)$
and $P_{2}(x,t=0)=P_{2}(1-x,t=0)$, then this property remains in time, 
$P_{1}(x,t)=P_{2}(1-x,t)$ and $P_{2}(x,t)=P_{2}(1-x,t)$ 
(the average opinion is conserved).
\begin{figure*}[!ht]
\begin{center}
\mbox{\includegraphics[clip,angle=270,width=.45 \textwidth]{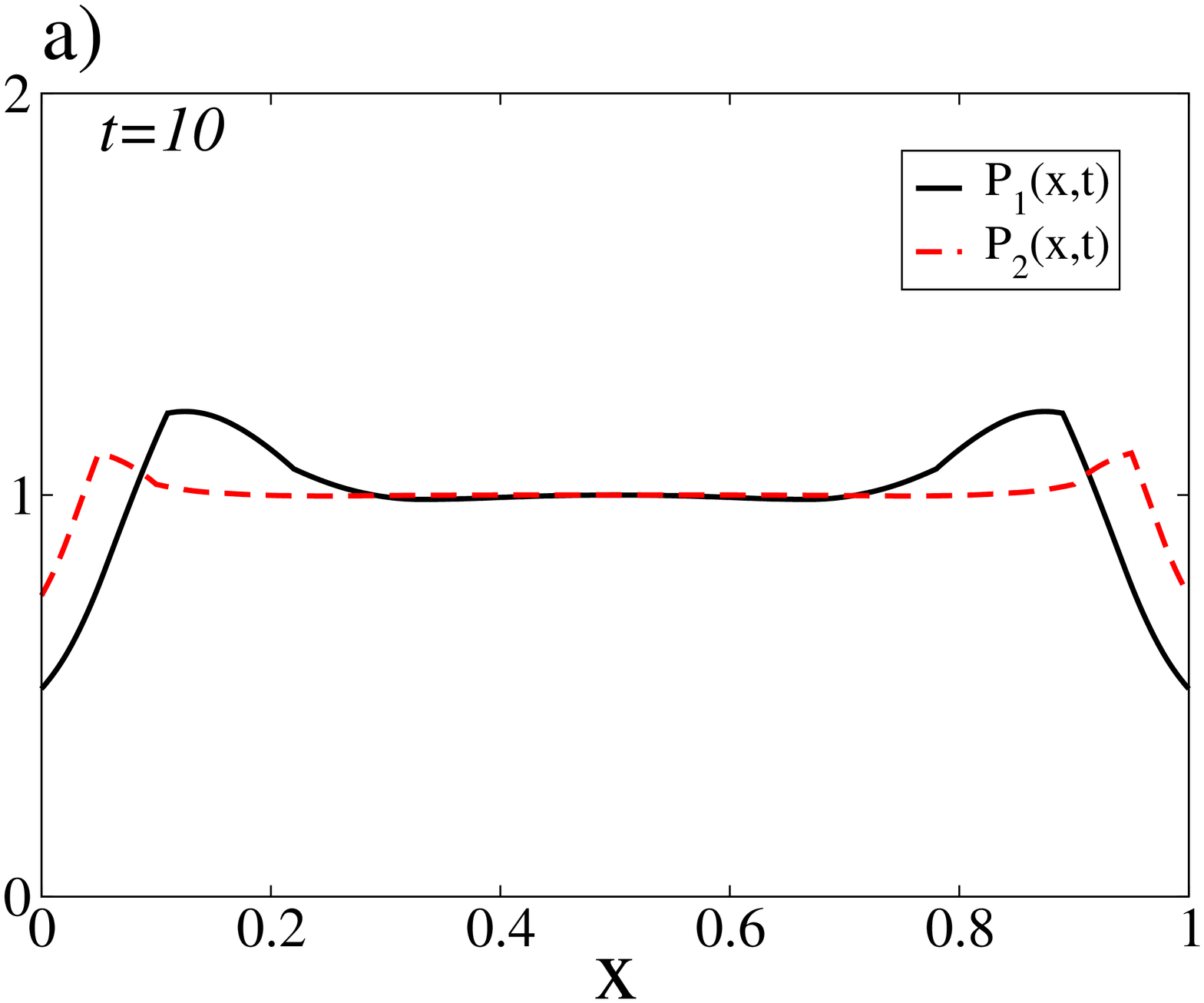}
\includegraphics[clip,angle=270,width=.45 \textwidth]{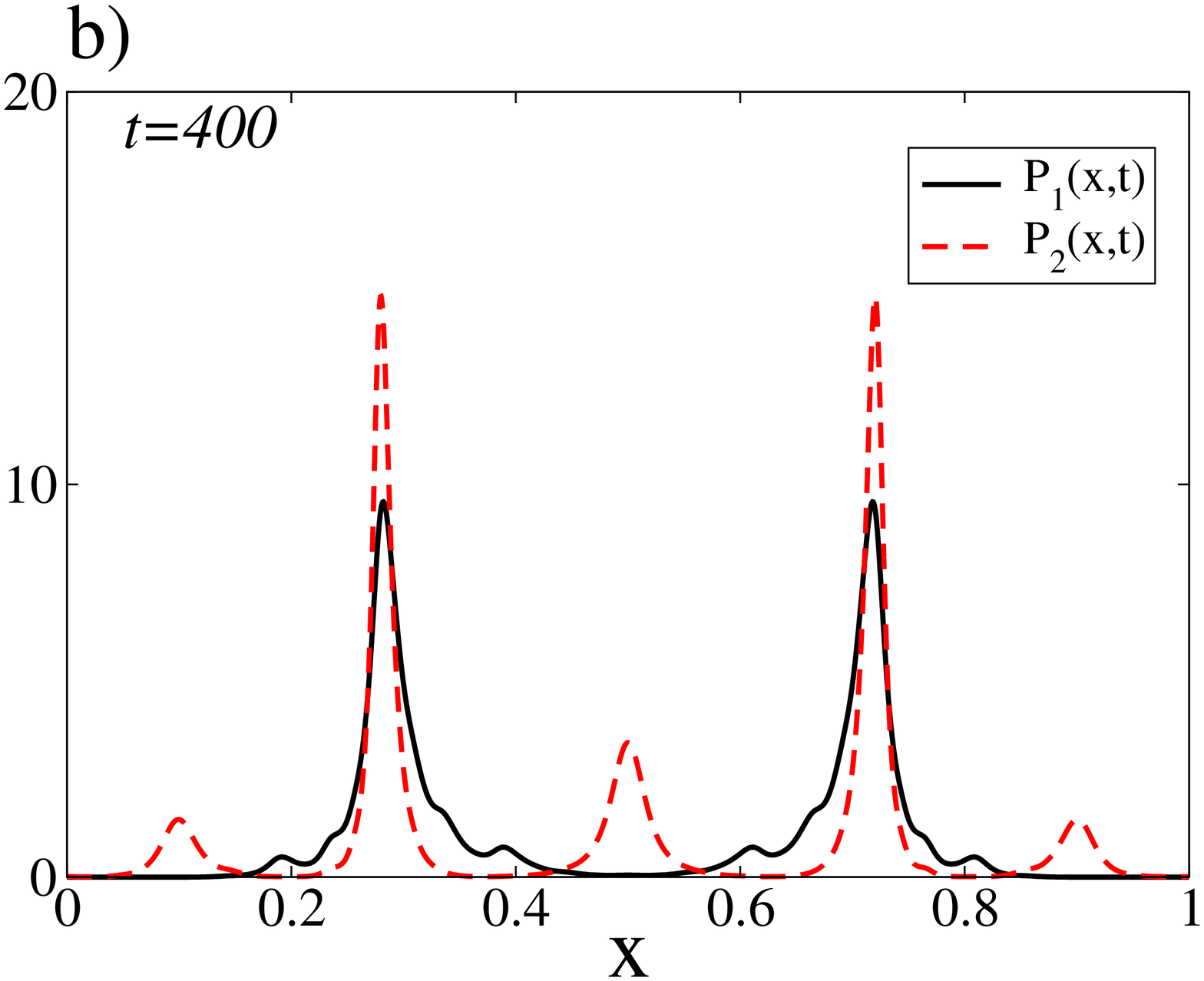}}
\hfill
\mbox{\includegraphics[clip,angle=270,width=.45 \textwidth]{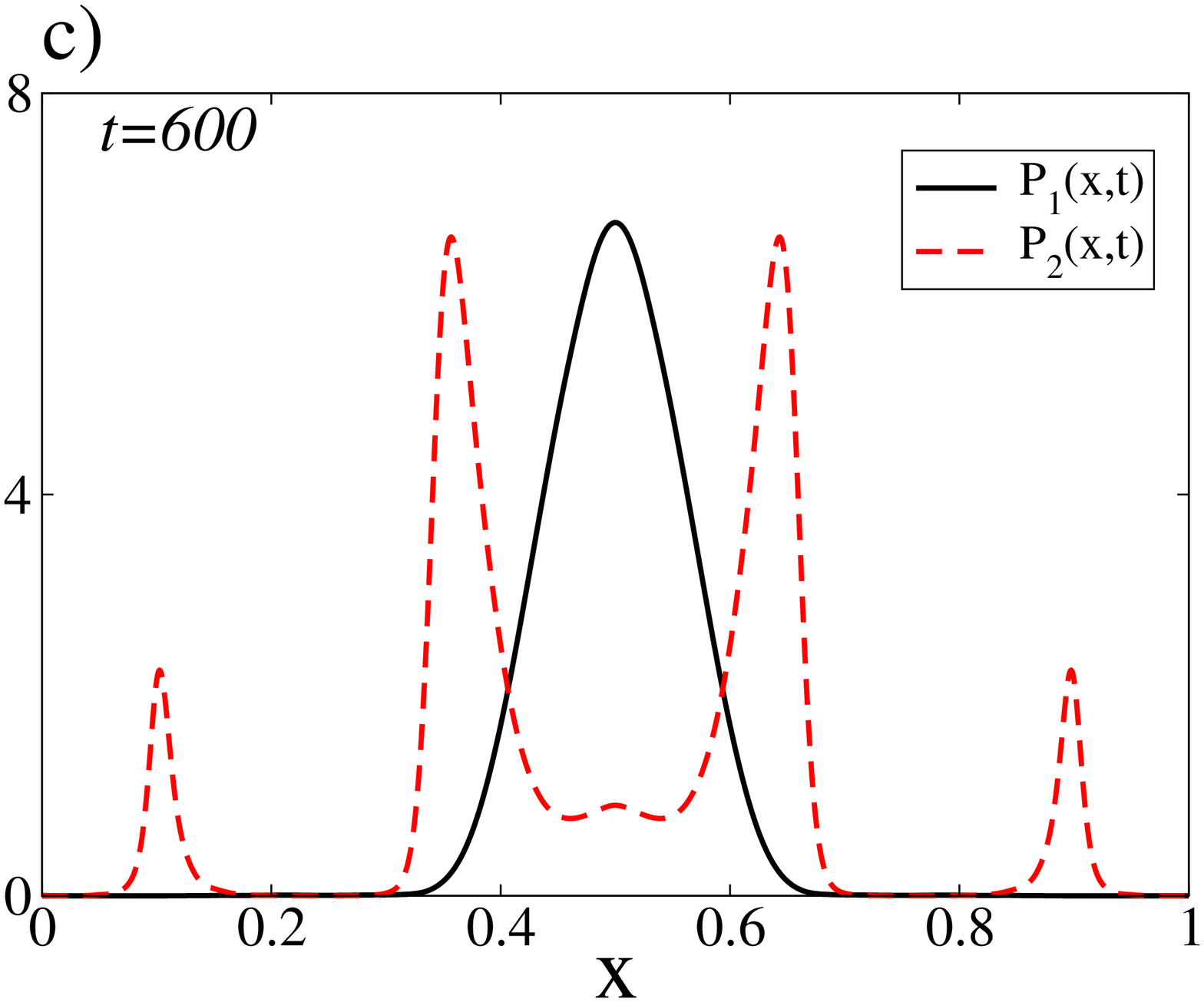}
\includegraphics[clip,angle=270,width=.45 \textwidth]{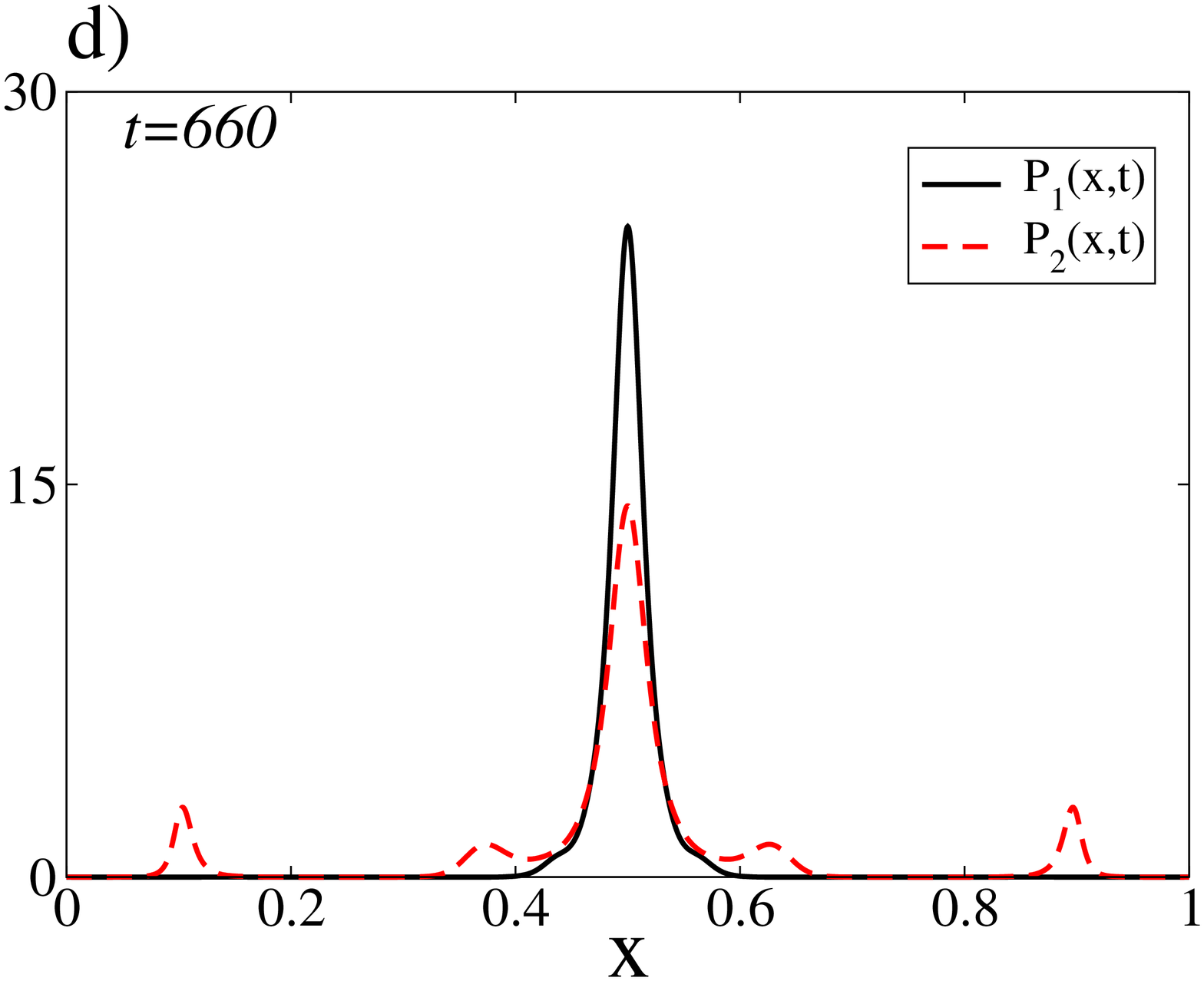}}
\caption[]{\footnotesize Probability distribution functions $P_{1}(x,t)$ and 
$P_{2}(x,t)$ for intermediate time steps starting
from a flat distribution with $\epsilon_{1}=0.22$ and $\epsilon_{2}=0.1$. We 
have used a fourth-order Runge-Kutta method with a 
time step $\Delta t=0.1$ for the time evolution, and the Simpson's rule for the 
integrals in $x$-space with a discretisation $\Delta x=1/M$, $M=2000$. 
The figures show the formation of consensus for confidence levels far below 
the critical point of the consensus transition of the homogeneous case 
($\epsilon_{1}=\epsilon_{2}\approx 0.27$) \cite{lorenz1,lorenz2}. The group of 
open-minded individuals are plotted using black-solid lines. For
the group of closed-minded individuals we use red-dashed lines.}
\label{fig10}
\end{center}
\end{figure*}
\begin{figure*}[!ht]
\begin{center}
\mbox{\includegraphics[clip,angle=270,width=.45 \textwidth]{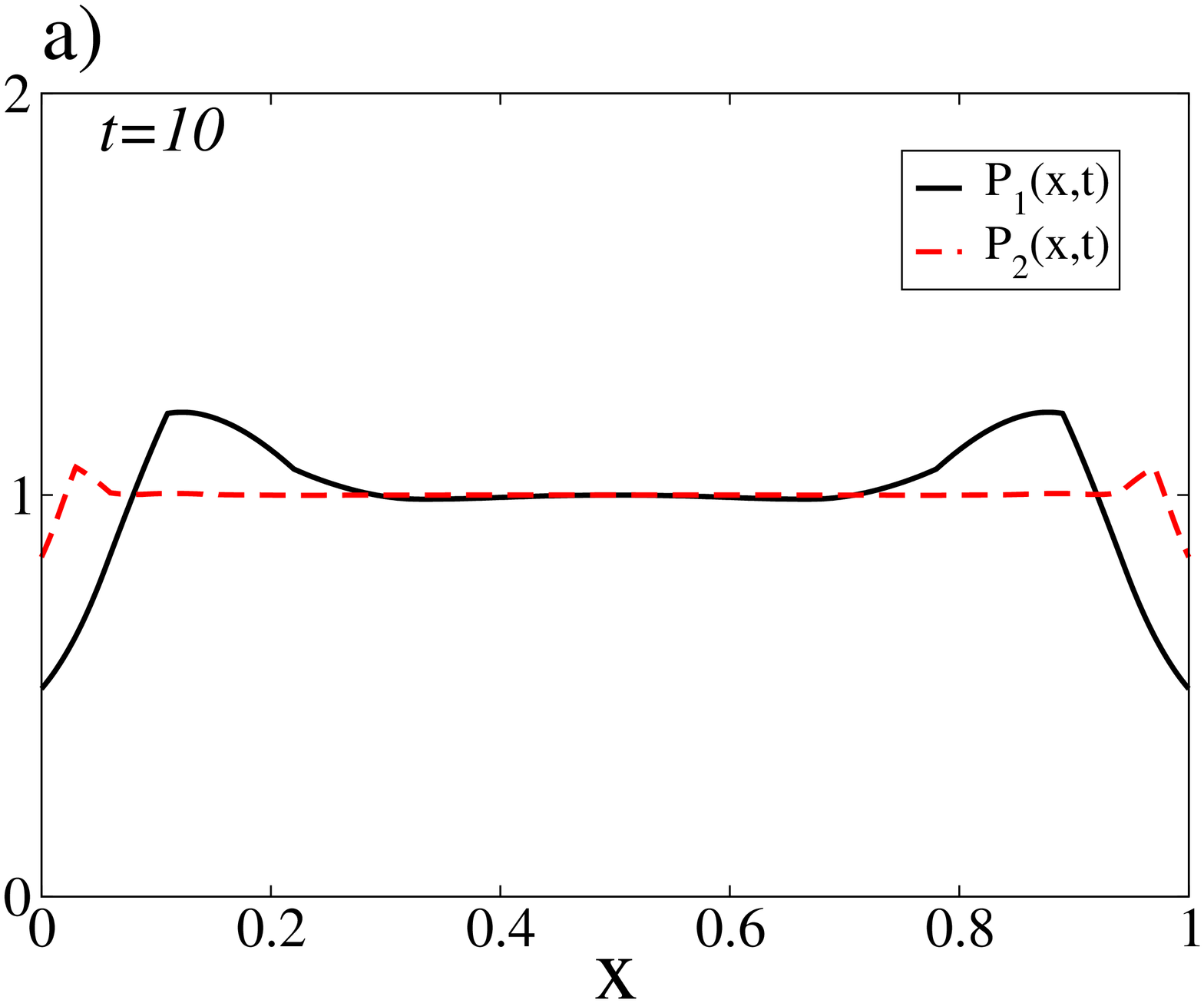}
\includegraphics[clip,angle=270,width=.45 \textwidth]{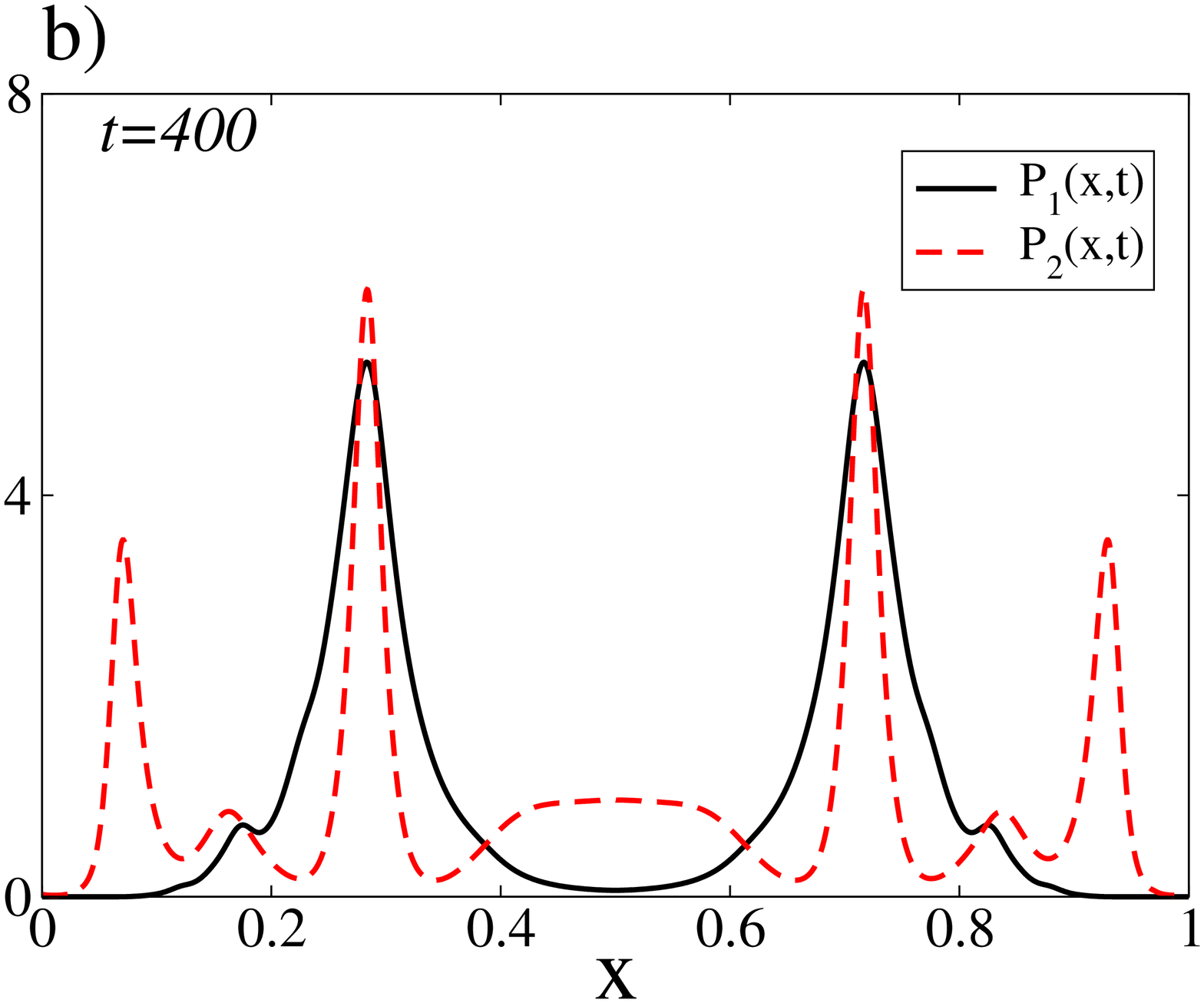}}
\hfill
\mbox{\includegraphics[clip,angle=270,width=.45 \textwidth]{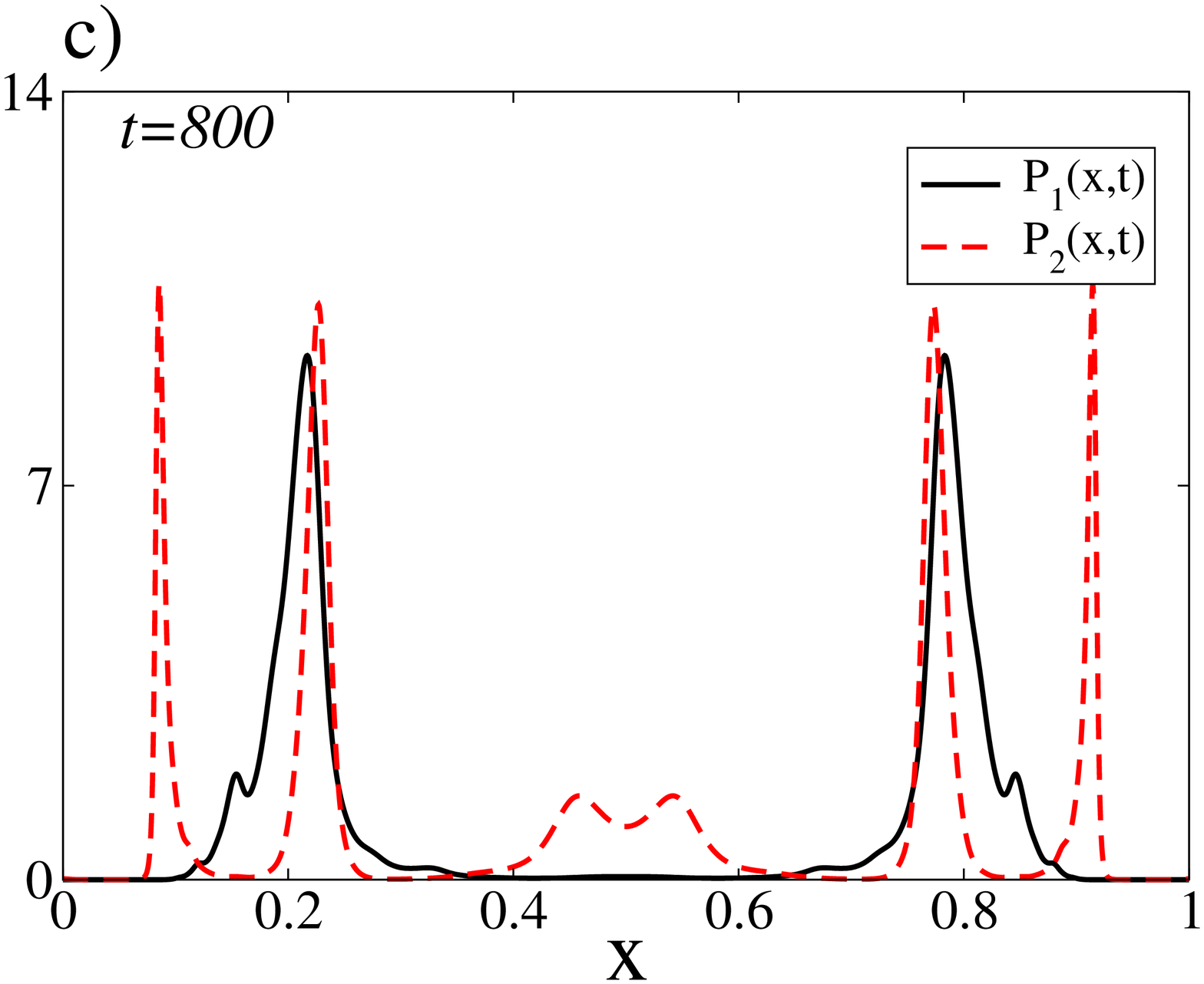}
\includegraphics[clip,angle=270,width=.45 \textwidth]{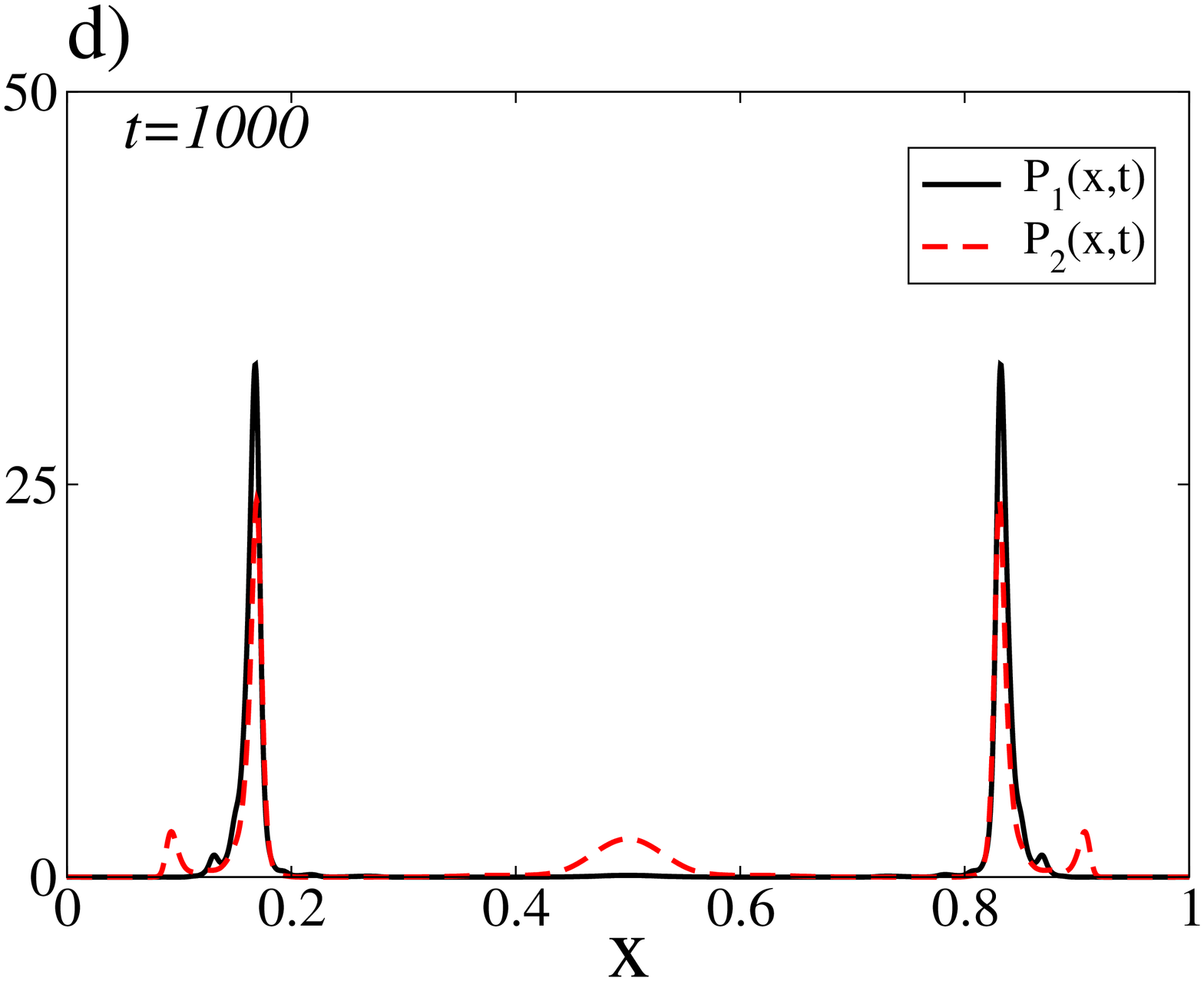}}
\caption[] {\footnotesize Probability distribution functions $P_{1}(x,t)$ and 
$P_{2}(x,t)$ for intermediate time steps starting
from a flat distribution with $\epsilon_{1}=0.22$ and $\epsilon_{2}=0.06$. We 
applied a fourth-order Runge-Kutta method with a 
time step $\Delta t=0.1$ for the time evolution, and the Simpson's rule for the 
integrals in $x$-space with a discretisation $\Delta x=1/M$, $M=2000$. 
The figures show how the opinions eventually split toward the two extremes. 
The group of open-minded individuals are plotted using black-solid lines. For
the group of closed-minded individuals we use red-dashed lines.}
\label{fig11}
\end{center}
\end{figure*}

In Fig.~\ref{fig10} and \ref{fig11}, we plot the time evolution of the probability distributions starting from
a uniform initial condition $P_{1}(x,t=0)=P_{2}(x,t=0)=1$ for $x \in[0.1]$. Fig.~\ref{fig10} shows the case for
$\epsilon_{1}=0.22$ and $\epsilon_{2}=0.1$. Initially, there is a group of closed-minded individuals in the center, 
coexisting with two big clusters composed by a mixture of open- and closed-minded individuals. Finally, the
open-minded individuals move to the center and pull the closed-minded individuals on
 both sides towards the center. This leads to the formation of a state of 
consensus (neglecting a small proportion of extremists). This result supports Monte Carlo simulations presented in Sec.~\ref{subsubsec:nomass},
which show that consensus can be achieved  as a consequence of mixing closed- and open-minded individual, even if 
both confidence levels are below  the critical point of the consensus transition of the homogeneous case
($\epsilon_{1}=\epsilon_{2} \approx 0.27$) \cite{lorenz1,lorenz2}. Similarly, Fig.~\ref{fig11}  shows a case for
$\epsilon_{1}=0.22$ and $\epsilon_{2}=0.06$. Initially, two big clusters of open 
and closed-minded individuals  are centered at $x \approx 0.75$ and $x \approx 0.25$. However, eventually, 
each big cluster starts to move towards $x=0$ or $x=1$ as a consequence of the 
interaction with the rest of 
closed-minded individuals whose opinions are closer to the extreme values.
 We notice that, starting from asymmetric or non-uniform initial distribution of opinions, could lead to the formation of 
 a single consensual state towards one extreme (as it occurs in Fig.~\ref{fig11} for each big cluster). Asymmetric perturbations of
 the initial state may appear naturally in Monte Carlo simulations because of the finite number of individuals, even if we
 start with uniform and essentially random distributions of opinions, this 
effect is more probable for smaller systems.

\end{document}